\documentclass[superscriptaddress,prb, reprint,longbibliography, 
 amsmath,amssymb,
 aps,
]{revtex4-2}
\usepackage{multirow}
\usepackage{graphicx}
\usepackage{dcolumn}
\usepackage{bm}
\usepackage{hyperref}
\hypersetup{colorlinks,allcolors=blue}
\begin{document}


\title{Magnetic phases of electron-doped infinite-layer Sr$_{1-x}$La$_x$CuO$_2$ \\ from first-principles density functional calculations}
\author{Alpin N. Tatan}
\email{alpinnovianus@g.ecc.u-tokyo.ac.jp}
\affiliation{Department of Physics, Graduate School of Science, The University of Tokyo, 7-3-1 Hongo, Bunkyo-ku, Tokyo, 113-0033, Japan}
\affiliation{The Institute for Solid State Physics, The University of Tokyo, 5-1-5 Kashiwanoha, Kashiwa, Chiba, 277-8581, Japan}
\author{Jun Haruyama}%
\affiliation{The Institute for Solid State Physics, The University of Tokyo, 5-1-5 Kashiwanoha, Kashiwa, Chiba, 277-8581, Japan}

\author{Osamu Sugino}

\affiliation{Department of Physics, Graduate School of Science, The University of Tokyo, 7-3-1 Hongo, Bunkyo-ku, Tokyo, 113-0033, Japan}
\affiliation{The Institute for Solid State Physics, The University of Tokyo, 5-1-5 Kashiwanoha, Kashiwa, Chiba, 277-8581, Japan}

\date{December 9, 2024}

\begin{abstract}
The magnetic phases of electron-doped infinite-layer Sr$_{1-x}$La$_x$CuO$_2$ are elucidated by first-principles density functional calculations. The antiferromagnetic parent state, metallic transition, as well as lattice evolution with doping and pressure are found to be consistent with experiments. The specific heat coefficient $\gamma$, magnetic exchange coupling $J$, as well as the density of states at Fermi level $N(0)$ of low-energy states with multiple magnetic configurations are investigated. We highlight a subset of such states in which we note an increase in $N(0)$ to suggest the interesting effects of magnetic fluctuations and La substitution on the electronic structure of this material. \\

\noindent This arXiv update is self-prepared to look similar to its published version: \\ A. N. Tatan. J. Haruyama, and O. Sugino. Phys. Rev. B \textbf{109}, 165134 (2024), which is available at DOI: \href{https://doi.org/10.1103/PhysRevB.109.165134}{10.1103/PhysRevB.109.165134}. All citations should refer to the published version.
\end{abstract}

\maketitle


\section{\label{sec:level1}Introduction}

\begin{figure}[b]
	\centering
		\includegraphics[width=1\linewidth]{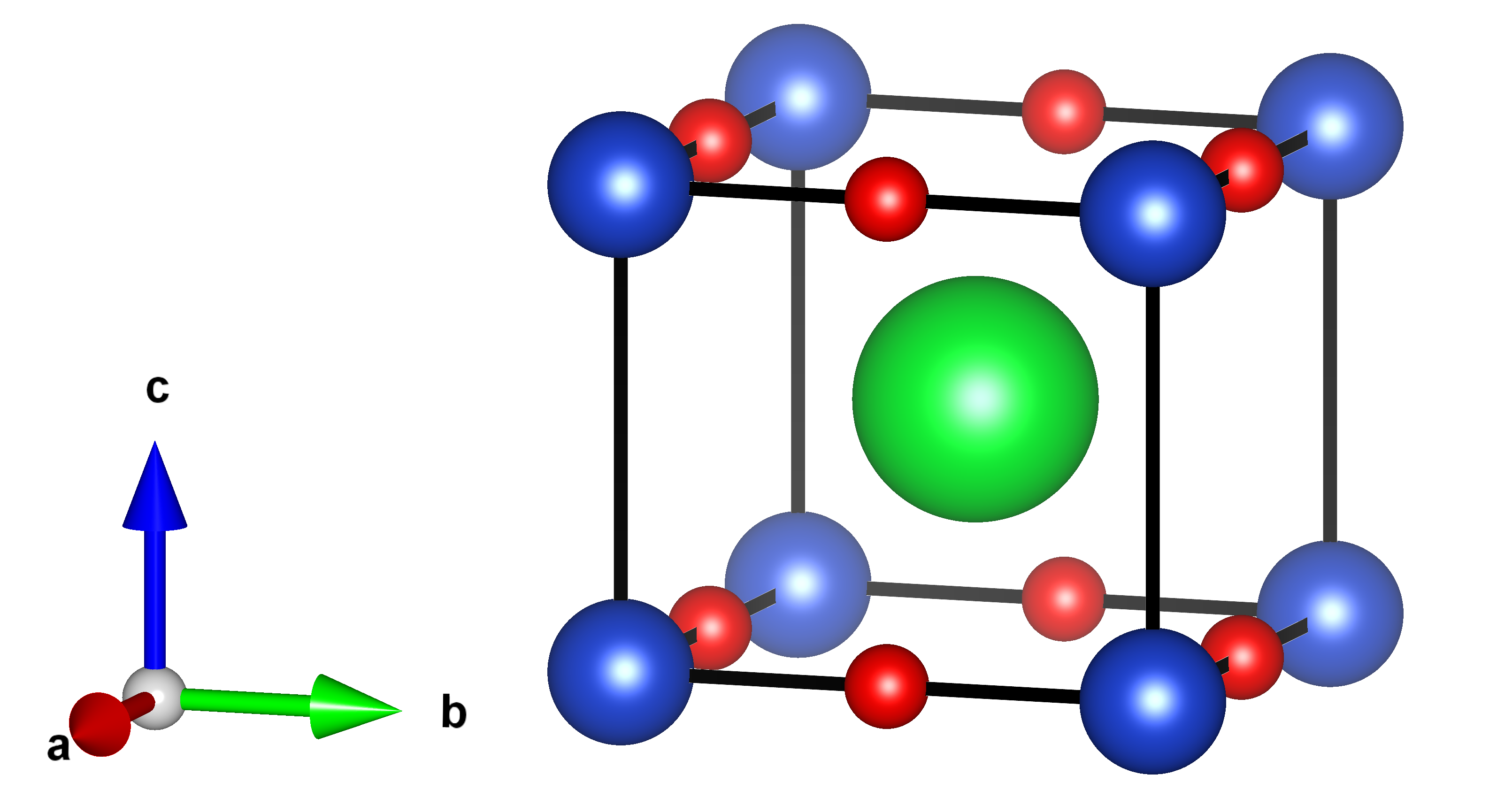}
	\caption{The unit cell of $\mathrm{Sr}_{1-x}\mathrm{La}_{x}\mathrm{CuO}_2$ (SLCO) with the atoms occupying Sr/La, O, and Cu sites represented by green, red, and blue spheres, respectively.} \label{FIG:1}
\end{figure}

Superconductivity can be achieved by doping cuprates with electrons from the antiferromagnetic (AFM) phase \cite{bednorz1986}. As this phase coexists with superconductivity at low doping levels, investigating its behavior in this region is beneficial for gaining insights on the latter \cite{Pickett1989, Damascelli2003, Armitage2010, FOURNIER2015, NAITO2016, Jovanovic2010}. Experiments on electron-doped cuprates have been mostly performed on $\mathrm{RE}_{2-x}\mathrm{Ce}_{x}\mathrm{CuO}_4$ compounds. Unfortunately, the magnetic rare-earth (RE) atoms can mask the information from the CuO$_2$ plane \cite{Jovanovic2010}. The $\mathrm{Sr}_{1-x}\mathrm{La}_{x}\mathrm{CuO}_2$ (SLCO, see Fig. \ref{FIG:1}) is an interesting alternative, having a simple infinite-layer tetragonal structure (space group \textit{P4/mmm}) free from the aforementioned issue. 
However, direct investigation of its AFM order from neutron scattering techniques is not yet possible due to the absence of single crystals. Nevertheless, its intriguing properties continue to be revealed from indirect measurements performed on polycrystalline and thin film samples \cite{Jorgensen1993,Jovanovic2021, Williams2005, Satoh2008, Harter2012, Harter2015}. For example, the presence of AFM order has been deduced from the nuclear magnetic resonance (NMR) wipeout effect and analysis of its linewidths \cite{Williams2005}. The angular magnetoresistance (AMR) has also been used as a similar probe \cite{Yu2007, Jovanovic2010, Jovanovic2021}. At low fields, an AFM alignment along the Cu-O-Cu direction prevailed at low $x$, whereas a state with mixed orientations could be found with increasing $x$ \cite{Jovanovic2010, Jovanovic2021}.

First-principles calculations based on density functional theory (DFT) are useful for complementing experimental investigations. In contrast to the local density or generalized gradient approximations (LDA/GGA), accurate descriptions of the AFM state have recently been achieved with the strongly constrained and appropriately normed (SCAN) functional \cite{Sun2015} for several parent compounds and hole-doped cuprates \cite{furness2018, Lane2018,zhang2020,Nokelainen2020,Tatan2022} without invoking the empirical Hubbard $U$ coefficient, which otherwise limits the predictive power of the theory. A comparison study of various functionals in La$_{2-x}$Sr$_{x}$CuO$_4$ \cite{pokharel2022sensitivity} also suggested SCAN as a preferred option to the costlier HSE06 hybrid functional \cite{HSE06}. Unlike SCAN, the hybrid functional was shown to inaccurately predict gapped states for both the parent and doped compounds, thus missing the insulator-to-metal transition \cite{pokharel2022sensitivity}. Interestingly, SCAN also predicts a landscape of competing AFM and stripe phases in the hole-doped cuprate, YBa$_2$Cu$_3$O$_7$ \cite{zhang2020}. The different Cu spin orders in these phases are expected to mix at finite temperatures, resulting in fluctuations in the pseudogap phase where superconductivity is believed to emerge from in this material \cite{zhang2020}. 

If the competing phases are universally responsible for the rich physics of cuprates, one naturally wonders if they manifest in some forms of exotic AFM phases in the electron-doped compounds given that there is still little experimental evidence for stripes formation in those materials \cite{Armitage2010}. In addition, a large part of pseudogap phenomena in electron-doped cuprates is believed to be tied to AFM correlations \cite{Armitage2010}. Granted, the description extracted from DFT calculations is limited to the normal state properties, and one should be cautious to not make quantitative predictions on superconductivity only from such results. Nevertheless, we believe it may be reasonable to consider the predicted normal-state behaviors as potential insights given the close-yet-subtle relationship between the two phases. For example, the evolution of experimentally measured superconducting transition temperature, $T_c$, from underdoped Bi$_2$Sr$_2$CuO$_{6+\delta}$ (Bi2201) and near optimally doped Bi$_2$Sr$_2$CaCu$_2$O$_{8+\delta}$ (Bi2212) cuprates with pressure can be qualitatively explained by comparison with the pressure dependence of the density of states at Fermi level, $N(0)$, from DFT calculations \cite{Deng2019}.

Following this line of thought, we hereby present several intriguing, low-energy AFM states in doped SLCO from total energy calculations performed in a similar manner to Ref. \cite{zhang2020} with some extensions. First, instead of considering all low-energy states altogether, we strive to identify only the states deemed favorable for a desired phenomenon (e.g., superconductivity). We consider $N(0)$ to be related to the likelihood of pair formation and treat it as a simple screening criteria based on its qualitative correlations described in literatures \cite{Deng2019,Sboychakov2008}. However, we remark that $N(0)$ should by no means be considered as a sole and final screening criterion and we hope future works will continue towards finer searches of the relevant states. Second, the low-energy states with higher $N(0)$ than the ground state are noted to be intermixtures of single-phase states (e.g., the ``G" or ``C" state in Ref. \cite{zhang2020}). This finding is confirmed across multiple doping and pressure levels, whereas Ref. \cite{zhang2020} considered only one doped composition (YBa$_2$Cu$_3$O$_7$) at ambient pressure. This observation regarding multiphase states is interesting since similar cases were found in manganites, where coexistence of magnetic phases has been used to explain the emergence of colossal magnetoresistance \cite{Dagotto2003}. 

Briefly, our results are presented as follows. First, we describe the ground state properties of the parent compound in Sec. \ref{SubsecParent}. Lattice structure evolution and metallic phase transition upon doping are presented here. Next, we summarize total energy calculations at $x=0$ and $x=0.125$ of structures with different spin alignments relative to the lattice vectors in Sec. \ref{SubsecMAE} for comparison with the AMR results \cite{Jovanovic2010,Jovanovic2021}. In Sec. \ref{LEMP}, we highlight several multiphase states in doped SLCO which lead to analysis of two features:
\begin{enumerate}
    \item the enhancement of $N(0)$ as a direct result of Cu magnetic fluctuations.
    \item a less seen effect of dopant substitution elucidated by the states consisting of both magnetic and nonmagnetic CuO$_2$ planes.
\end{enumerate} 
These multiphase states are benchmarked with estimation of relevant parameters, e.g., the magnetic exchange coupling, $J$, and specific heat coefficient, $\gamma$, which agree with the expected values from experiments. Finally, we show how the ground state and the multiphase states evolve under application of external pressures in Sec. \ref{SubsecPress} to conclude this brief study.

\section{Computational Methods}
\begin{figure}[b]
	\centering
		\includegraphics[width=1\linewidth]{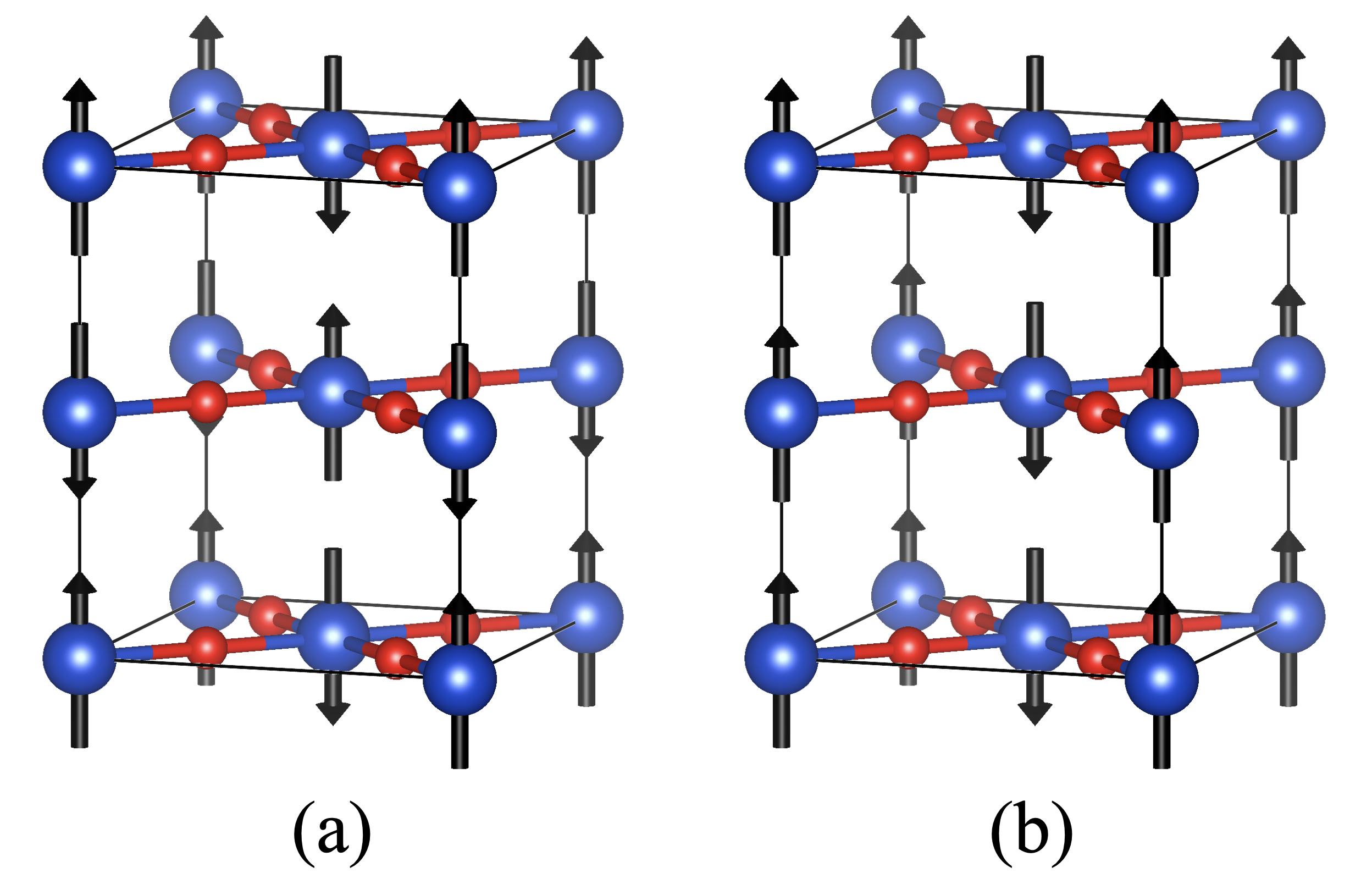}
	\caption{Schematics of the (a) ``G" and (b) ``C" AFM configurations in a $\sqrt{2}\times\sqrt{2}\times 2$ supercell. The AF moments are drawn to align normal to the CuO$_2$ plane for simplicity. Blue and red spheres represent Cu and O atoms. Sr atoms have been omitted for clarity.} \label{FIG:G_C_order}
\end{figure}
Generally, we follow the preceding SCAN studies for cuprates \cite{furness2018,Lane2018,zhang2020,Nokelainen2020,Tatan2022}. Calculations are performed with the pseudopotential projector augmented-wave (PAW) method \cite{Kresse1999} implemented in the Vienna ab initio simulation package (VASP) \cite{Kresse1993,Kresse1996} with an energy cutoff of 550 eV for the plane-wave basis set. Exchange–correlation effects are treated using the SCAN meta-GGA scheme \cite{Sun2015}. Structures are optimized using blocked-Davidson and preconditioned conjugate gradient algorithms with an atomic force threshold of 0.01 eV/\textup{\AA} and a total energy threshold of $10^{-5}$ eV. $\Gamma$-centered k-point mesh with spacing of 0.12 \AA$^{-1}$ is used for structure optimizations. We tighten the energy threshold to $10^{-8}$ eV and utilize denser k-point meshes (0.08 \AA$^{-1}$ or less) in total energy and density of states calculations using the tetrahedron method with Bl\"ochl corrections \cite{Blochl1994}. Most of the calculations are performed in collinear mode. For comparison with the AMR results \cite{Jovanovic2010,Jovanovic2021}, noncollinear calculations that consider spin-orbit coupling effects are performed at $x=0$ and $x=0.125$ to compute the magnetic anisotropy energy (MAE), defined as: 
\begin{equation}
    \mathrm{MAE}_{xyz} = E_{xyz} - E_c
    \label{eq1}
\end{equation}
where $E_{xyz}$ and $ E_c$ are, respectively, the total energies calculated with magnetic moments on copper sites aligned along a direction shown by $[xyz]$ and along the $c$-axis in the coordinate system depicted in Fig. \ref{FIG:1}. To match with Refs. \cite{Jovanovic2010,Jovanovic2021}, we consider MAE calculations for AFM alignments along either side of the CuO$_2$ plane (Cu-O-Cu line) and along the face diagonal of the CuO$_2$ plane (Cu-Cu line), respectively. Electron doping is achieved by substituting trivalent La on divalent Sr sites using the supercell method. 

\begin{figure}[h]
    \centering
    \includegraphics[width=1\linewidth]{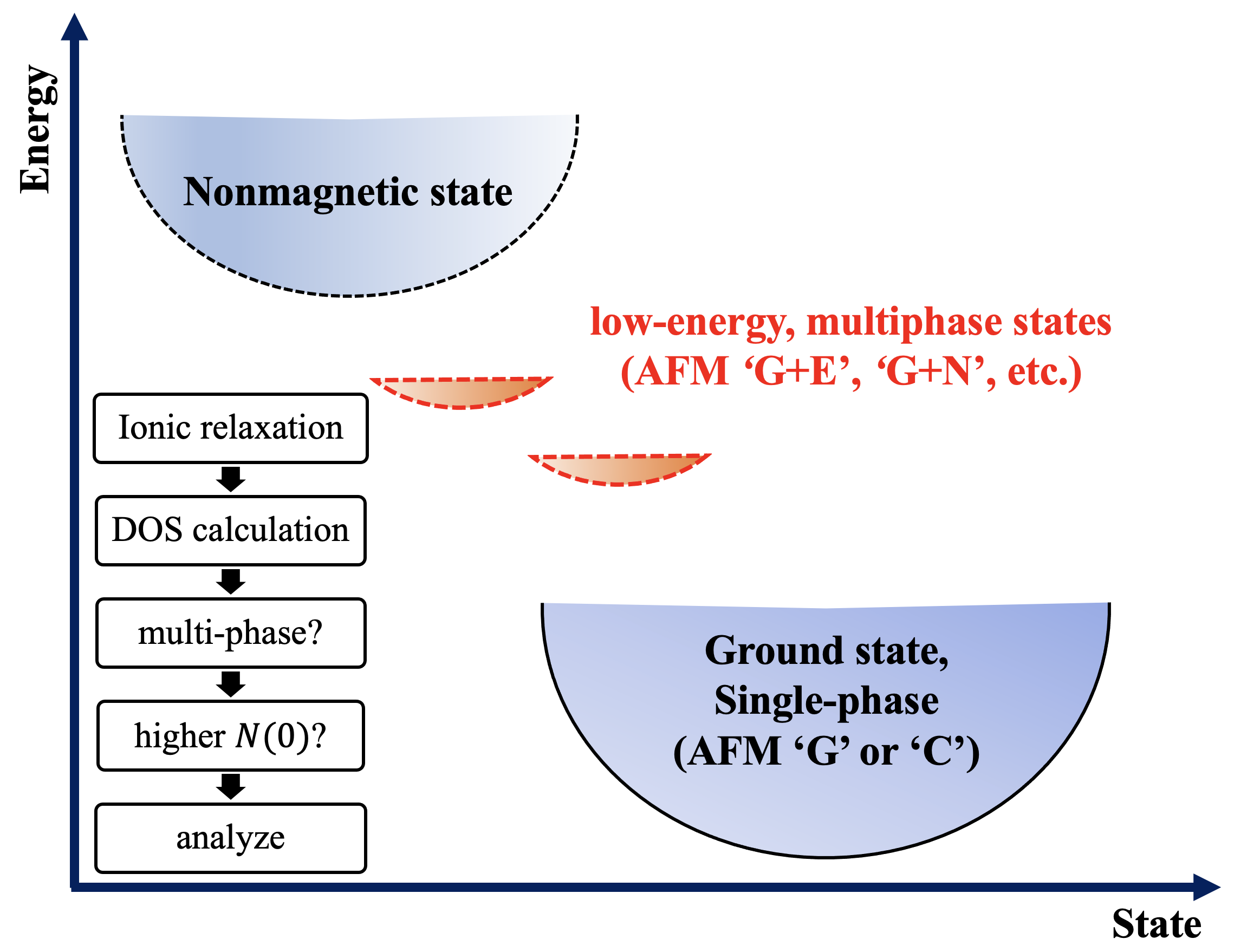}
    \caption{A schematic energy diagram of our search. The AFM G, C, and nonmagnetic states are robust. Most of our calculations from various starting conditions converge into these single-phase states, which we illustrate with deep arcs (large curvature). The ground state is drawn with solid curves to indicate its position as the most stable state. The multiphase states are low-energy states that our calculations may end up converging into, given suitable initial parameters. These are drawn in shallow arcs (small curvature). We search for such local minima with increased DOS at Fermi level, $N(0)$.} 
	\label{FIG:flowchart}
\end{figure}
We perform our calculations with several ferromagnetic and antiferromagnetic states described in Ref. \cite{Wollan1955} as ``A"-type, ``B"-type magnetic order, etc. This nomenclature has also featured in recent works across different materials \cite{zhang2020,zhang2021nickelate}. We also compute the nonmagnetic ``N" state as an additional energy reference, given the robustness of this state in self-consistent calculations. We attempt these calculations in supercells in various sizes ($\sqrt{2}\times\sqrt{2}\times2, 2\times 2\times 5, 4\times 4\times 2$, etc.) to moderately balance the coverage of initial states and the computational cost. We emphasize that our collinear calculation for the magnetic states is not a constrained magnetic calculation (which would otherwise be a noncollinear calculation and require a penalty functional) \cite{Ma_constrained}. Thus, despite the different initial states, most of our calculations converge into the lowest energy G and C configurations [Figs. \ref{FIG:G_C_order}(a) and \ref{FIG:G_C_order}(b)]. They have similar energies and the configurations differ only in whether interplanar AFM order is present. These states are also the AFM states with lowest energy for Y-based cuprates in Ref. \cite{zhang2020}. Moreover, we do expect a number of local minima beyond these single-phase magnetic states to which our self-consistent optimizations can converge at suitable conditions. In addition to finding the ground state with the robust preconditioned conjugate gradient algorithm, local minima can also be found using the blocked-Davidson scheme as long as convergence is reached. As the latter scheme is not robust for finding the ground state, we use it to conveniently find nearby local minima in standard self-consistent calculations. We utilize this method in a screening procedure as illustrated in Fig. \ref{FIG:flowchart} where we look for interesting changes of the electronic structure (e.g., a higher $N(0)$ value) in these local minima \cite{Dagotto2003,dagotto2003book}. 

\section{Results and Discussions}
\subsection{\label{SubsecParent}Doping the parent compound: metallic phase transition and lattice parameter evolution}

The optimized lattice parameters of the parent compound take the value of $a = 3.910$ \AA, $c = 3.444$ \AA, which are less than 1\% away from the experimental values ($a = 3.927$ \AA, $c = 3.435$ \AA) in Ref. \cite{kikkawa1992}. These values are insensitive to the choice of magnetic configurations. The ground state is the G configuration, with the C and N configurations lying 3.28 and 59.11 meV above it. The magnetic moment at copper site is calculated by integrating the magnetic moment within a PAW sphere of radius 1.164 \AA,  as set by the pseudopotential for Cu. For the parent compound, this is approximately $\pm 0.47 \mu_\mathrm{B}$ per atom. 

\begin{figure}[h]
    \centering
    \includegraphics[width=1\linewidth]{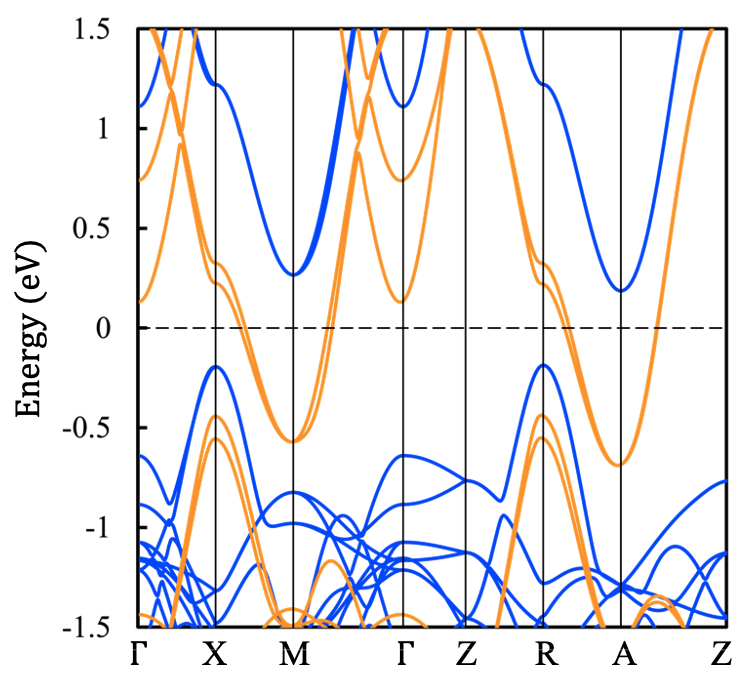}
    \caption{The plotted band structure for SLCO with $x=0$ (blue) and $x=0.25$ (orange) \cite{Ganose2018}. Only the spin-up bands are shown. The Fermi level is set to zero. A metallic phase transition with electron doping character is achieved.}
    \label{fig:rgb_mit}
\end{figure}

\begin{figure}[h]
    \centering
    \includegraphics[width=0.9\linewidth]{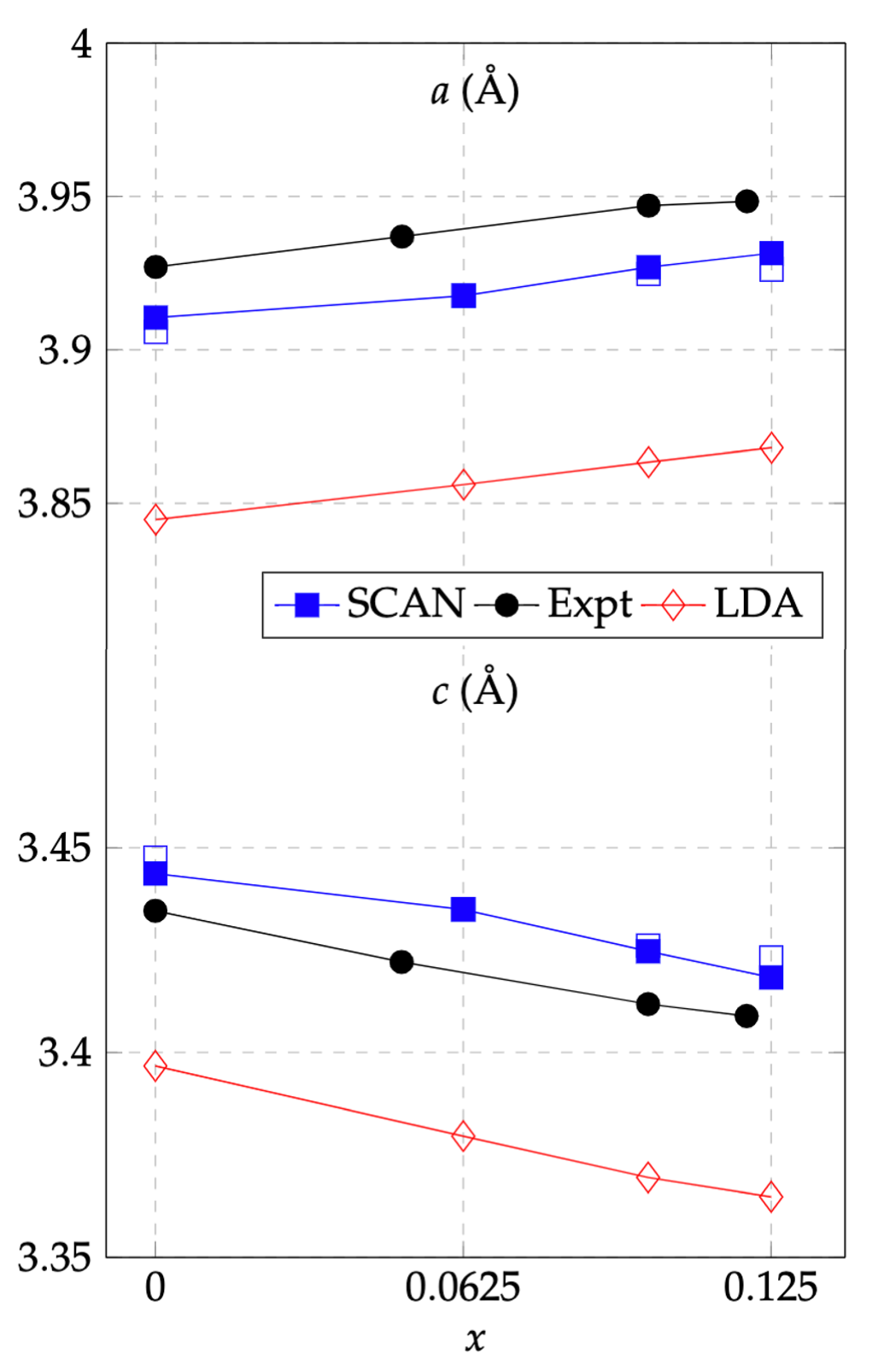}
    \caption{The lattice parameters versus $\mathrm{La}$ concentration $x$. The open (closed) symbols denote DFT values without (with) magnetic order. Experimental values are from Ref. \cite{kikkawa1992}.}
    \label{fig:latticeparam}
\end{figure}

The band structure of the parent compound is plotted in Fig. \ref{fig:rgb_mit} (blue lines). The low-energy bands are primarily of Cu $3d$ and O $2p$ orbital characters, as commonly found in cuprate compounds \cite{FOURNIER2015,Lane2018}. A band gap opens when AFM order exists. For the ground state with G configuration, an indirect band gap of approximately $0.35$ eV exists between the R and A points of the first Brillouin zone of this $\sqrt{2}\times\sqrt{2}\times 2$ cell. The energy difference between the highest valence band and the lowest conduction band at R and A points are 1.41 and 1.47 eV, respectively. The smallest direct gap of approximately 1.1 eV exists at the M point. Interestingly, when the structure takes the C configuration, the energy difference between the highest valence band at the X point and the lowest conduction band at the M point decreases to approximately $0.2$ eV and becomes the effective indirect gap. Since the C configuration only has in-plane AFM order, this suggests the importance of interplanar interactions for the electronic structure of this material. Metallic phase transition is achieved by doping the parent structure, as exemplified for $x=0.25$ (orange lines in Fig. \ref{fig:rgb_mit}) where 1 La has been substituted for Sr in a $\sqrt{2} \times \sqrt{2} \times 2$ supercell. The band gap closes and the Fermi level moves to higher energy, indicative of a transition enabled by electron doping.

Subsequently, we expand the supercells to simulate lower La concentrations. The relaxed lattice parameters evolve with $x$ as shown in Fig. \ref{fig:latticeparam}, where the doped compounds are metallic. The value of the lattice parameters and their $x$-dependence are in good agreement with experiment \cite{kikkawa1992}. The shrinking $c$ parameter can be explained by the substitution of Sr$^{2+}$ (ionic radius 1.26 \AA) by the trivalent La$^{3+}$ (ionic radius 1.16 \AA) in the doped compounds while the $a$ parameter expansion has to do with electrons being doped into the antibonding $\sigma_{x^2-y^2}$ orbitals within the CuO$_2$ plane \cite{ER1991206}. Overall, the cell volume per formula unit increases slightly by 0.1745 \AA$^3$ (0.33 $\%$) from $x=0$ to $x=0.125$. The LDA, on the other hand, significantly underestimates the lattice parameters, which may be ascribed to the well-known problem of `overbinding'. This better agreement for SCAN is not attributed to magnetism, as our SCAN structures yield similar values for both magnetic and nonmagnetic cells. Instead, this showcases how the LDA and SCAN functionals treat bonds between atoms differently; indeed, accurate description of interatomic bonds is one of the focus areas of SCAN in its construction \cite{Sun2015}.

\subsection{\label{SubsecMAE}Magnetic anisotropy energy}

\begin{figure}[h]
    \centering
    \includegraphics[width=1\linewidth]{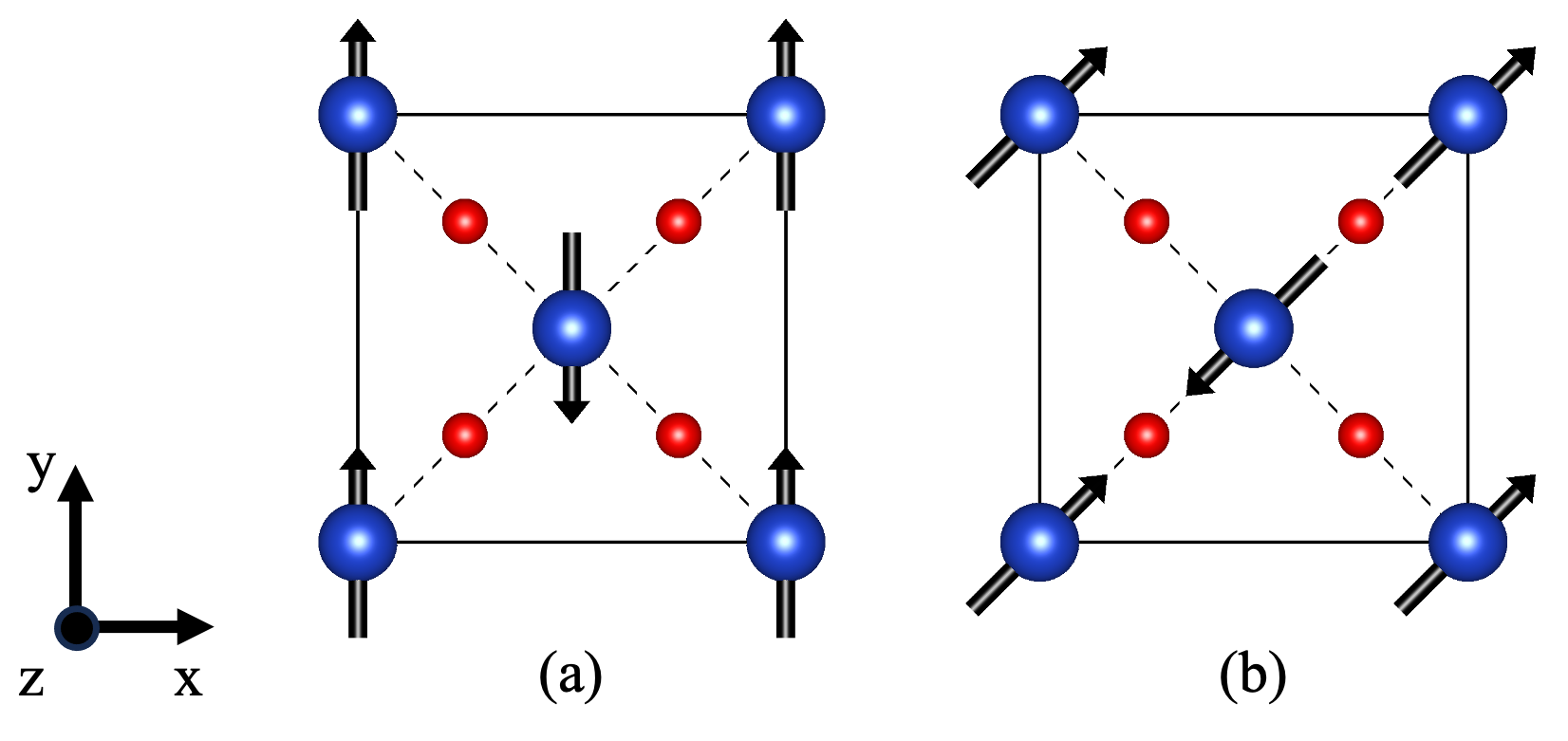}
    \caption{The top view of AFM supercells for the magnetic anisotropy energy study. Blue and red spheres represent Cu and O atoms. Black vectors represent the sign of magnetic moment at Cu sites. Dashed lines indicate the direction along the Cu-O-Cu bond. Structures with AFM order aligned within the CuO$_2$ plane along the (a) Cu-Cu and (b) Cu-O-Cu direction are shown.}
    \label{fig:MAEorder}
\end{figure}

Several noncollinear magnetism and spin-orbit coupling calculations are performed at $x=0$ and $x=0.125$ to determine the preferential AF alignment direction of Cu, in comparison with the $x=0.12$ samples measured in the AMR experiments \cite{Jovanovic2010,Jovanovic2021}. We assume that these experimental results at low fields also reflect the most favorable alignments in the zero-field limit, and we verify them with total energy calculations in this work. We use Eq. \eqref{eq1} for the respective alignments (Fig. \ref{fig:MAEorder}) and summarize the results in Table \ref{tblMAE}.

\begin{table}[h]
\caption{\label{tblMAE}%
Magnetic anisotropy energy (MAE) per Cu atom of $\mathrm{Sr}_{1-x}\mathrm{La}_{x}\mathrm{CuO}_2$ for the AF in-plane alignments along the Cu-Cu [Fig. \ref{fig:MAEorder}(a)] and Cu-O-Cu [Fig. \ref{fig:MAEorder}(b)] directions. Values are in meV and relative to the total energy for AF alignments along the $c$-axis (normal to the CuO$_2$ plane) }
\begin{ruledtabular}
\begin{tabular}{lcr}
Alignment  & $x=0$ &$x=0.125$ \\
\hline
Cu-O-Cu  &  -0.022 & +0.017 \\
Cu-Cu &  -0.002 &  +0.020 \\
\end{tabular}
\end{ruledtabular}
\end{table}

The MAE in cuprates is known to have small values. For example, the Li-based cuprate, Li$_2$CuO$_2$ has values between 0.045-0.078 meV per Cu atom, determined from  magnetization jump and neutron scattering experiments \cite{Boehm_1998,Mertz2005}. Hence, our computed MAE value $\approx 0.02$ meV per Cu atom is quite reasonable. The results indicate a preferential in-plane spin alignment along the Cu-O-Cu direction over the other directions for $x=0$, and along the $c$-axis at $x=0.125$. The small MAE suggests that it should be fairly common to obtain samples with either alignment, subject to synthesis conditions. The apparent degeneracy of in-plane orientations (along the Cu-Cu and Cu-O-Cu directions) suggests a less rigid AFM order in the doped compounds, which agrees with the AMR reports of Refs. \cite{Jovanovic2010, Jovanovic2021}.

\subsection{\label{LEMP}Low-energy magnetic phases in doped SLCO}

\begin{figure}[b]
    \centering
    \includegraphics[width=0.92\linewidth]{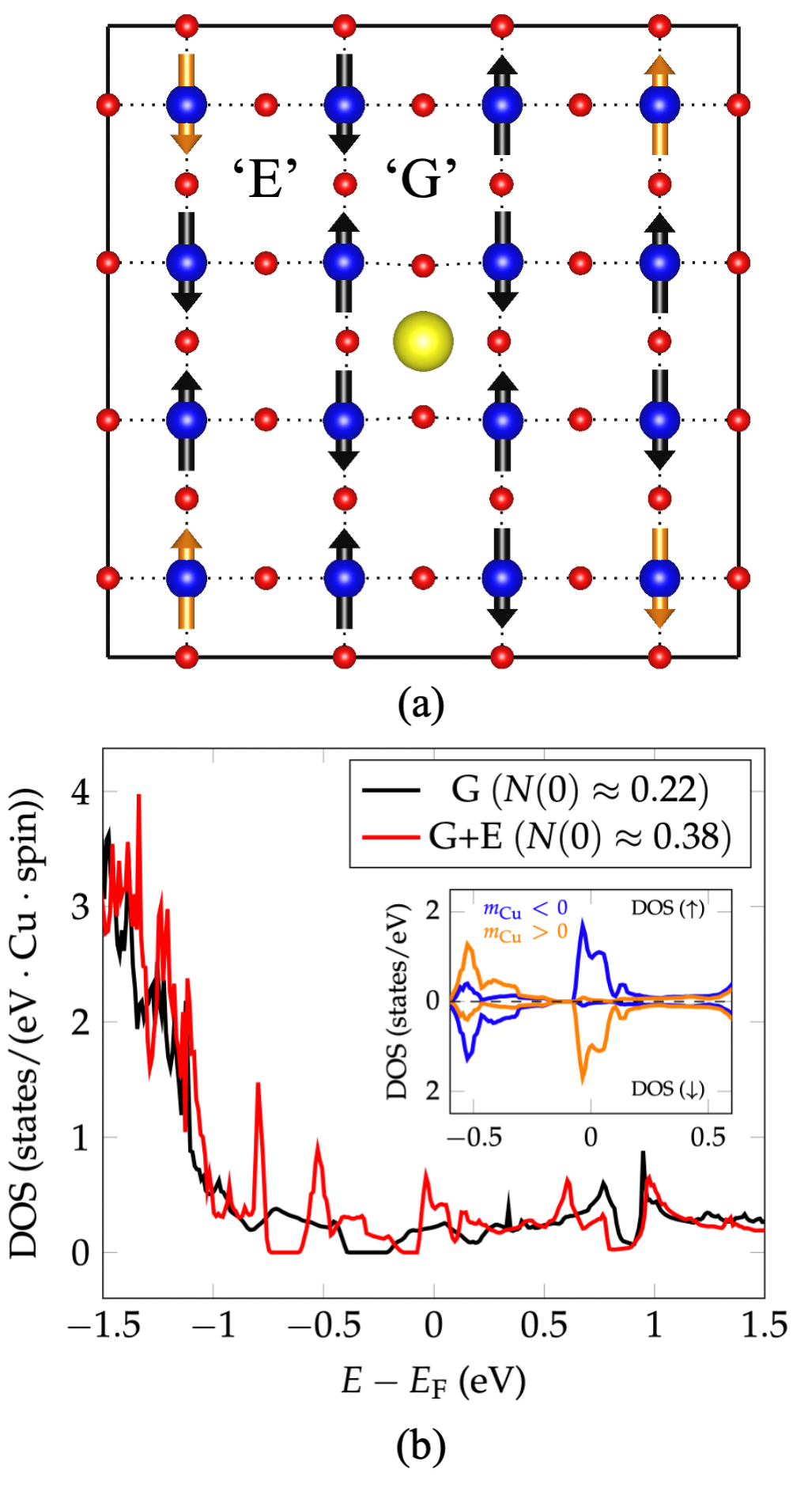}
    \caption{(a) The top view of a $4\times4\times2$ supercell in the G+E state for SLCO with $x=0.0625$. Yellow, red, and blue spheres represent
La, O, and Cu atoms. Sr atoms have been omitted for clarity. The magnetic moments are drawn along the CuO$_2$ plane for simplicity. The orange vectors illustrate the magnetic moments at corner sites, for which an E-type magnetic order \cite{Wollan1955} can be observed. Dotted lines mark the Cu-O bonds. (b) the DOS normalized per Cu atom of G and G+E states. The Fermi level is set to zero. Only the spin-up DOS is shown. Inset: The DOS from single Cu atom located at the corner sites in the G+E state. Upper and lower halves of the inset illustrate the spin-up and spin-down DOS, respectively. The orange (blue) curve marks the DOS from sites with positive (negative) magnetic moment $m$.}
    \label{fig:0_0625}
\end{figure}

The total energies of relaxed structures relative to the magnetic G state are summarized in Table \ref{tab:energy} with normalized values per Cu atom. The G state is always more stable than the nonmagnetic N state, but electron doping progressively reduces their energy difference. Additionally, multiphase states, (e.g., G + N state), with energies lying between the nonmagnetic and single-phase G state can be found in the doped compounds. The C state is generally very close in energy to the G state, considering the resolution of these calculations. Because the C state does not improve the values of $N(0)$ with respect to the G state, we use the latter as reference for analysis of the multiphase states. As mentioned in the Introduction, multiphase states that do not satisfy the DOS and total energy criteria: $N(0) > N_{\mathrm{G}}(0); ~ E_{\mathrm{G}} < E < E_{\mathrm{N}} $ have been omitted from Table \ref{tab:energy}.

\begin{table}[h]
    \caption{Total energy $E$ per Cu atom of $\mathrm{Sr}_{1-x}\mathrm{La}_{x}\mathrm{CuO}_2$ relative to the magnetic G state in units of $\mathrm{meV}$. The values for each $x$ are computed on $\sqrt{2}\times \sqrt{2} \times 2$, $4\times 4\times 2$, $2\times 2\times 5$, and $\sqrt{2}\times \sqrt{2} \times 4$ supercells, respectively. The nomenclature (G, C, E) follows Ref. \cite{Wollan1955}. Only multiphase states with increased $N(0)$ are shown. At $x=0.1$ and $x=0.125$, the G+N state with nonmagnetic planes located closer (further) from the dopant site is denoted by normal (bold) fonts.}
    \label{tab:energy}
    \centering
\begin{ruledtabular}
    \begin{tabular}{c c c c c}
      Config. & $x=0$ & $x=0.0625$ &$x =0.1$ & $x=0.125$\\
      \hline
    N&  +59.11 & +50.54 & +47.30 & +37.85\\
     C &  +3.28 & -0.09 & -0.13 & -0.19\\
     Multiphase &  &   & +10.64 &  +14.62\\
      &  & & \footnotesize{(G+C+N)} &  \footnotesize{(G+N)}\\
      &  &+31.77    & +14.03 &  +22.79 \\
      &  &\footnotesize{(G+E)}  & \footnotesize{(G+N)} &  \textbf{\footnotesize{(G+N)}} \\
      &  &  & +31.93 &  \\
      &  &  & \textbf{\footnotesize{(G+N)}} & \\
    \end{tabular}
\end{ruledtabular}
\end{table}

\subsubsection{Magnetic moment fluctuation and the DOS at Fermi level: a study of the `G+E' state at $x=0.0625$}

In Fig. \ref{fig:0_0625}(a), we highlight the low-energy G+E state found at $x = 0.0625$. This multiphase state is similar to the single-phase G state, except that the magnetic moments at the corner Cu sites are sign reversed (orange vectors). The size of magnetic moments at copper sites is also less uniformly distributed in the cell: the G state has magnetic moments at copper site of $ \pm 0.396 - 0.412\mu_\mathrm{B}$ per atom while the G+E state has magnetic moments at copper sites of $\pm 0.309 - 0.394\mu_\mathrm{B}$ per atom.

We show the density of states from the G and G + E state in Fig. \ref{fig:0_0625}(b). $N(0)$ is enhanced by almost 73\% in the G+E state relative to the G state. Moreover, a peak exists close to $E_\mathrm{F}$ with $N_\mathrm{G+E}(-0.035 ~\mathrm{eV}) \approx 3N_\mathrm{G}(0)$. According to atom-resolved DOS, this increase originates from the corner Cu sites. We have plotted the DOS for both spins from a single Cu atom at these sites in the inset of Fig. \ref{fig:0_0625}(b). The Cu site with a negative magnetic moment value (blue curve) has a peak for the spin-up DOS (upper half of the inset figure) at the Fermi level and vice versa (orange curve). Considering its low-energy nature, this result suggests an enhancement mechanism for $N(0)$ arising from magnetic fluctuations. At finite temperatures, the single-phase ground state would mix with the low-energy states and then the Cu states will be shifted towards the Fermi level with an effect of enhancing $N(0)$. This may help to improve the conditions for Cooper pair formation.

Here, the energy difference between G + E and G states is used to estimate the magnetic exchange coupling parameter $J$. We consider the total energy of a periodic supercell containing $N$ magnetic atoms to consist of a nonmagnetic part  $E_0$ and magnetic part $E_\mathrm{mag}$:
\begin{equation}
    E(N \mathrm{~magnetic~atoms}) = E_0 +E_\mathrm{mag}
\end{equation}
The following approximation based on spin-$\frac{1}{2}$ nearest-neighbor (NN) 2D Ising model is used for $E_\mathrm{mag}$:
\begin{align}
    E_\mathrm{mag} &= \mathrm{No.~ of~ FM~ NN~ pairs} \times \left(\frac{J}{4}\right) \nonumber \\ 
    &+ \mathrm{No.~ of~ AFM~ NN~ pairs} \times \left(\frac{-J}{4}\right)
\end{align}
The energy difference of these states can be computed as:
\begin{align}
    \Delta E_\mathrm{G+E,G} &= \Delta E_0 + \Delta E_\mathrm{mag} \nonumber \\ &\approx \frac{J}{4} \times \Bigg\{  \Delta \left[\mathrm{No.~ of~ FM~ NN~pairs}\right]  \nonumber \\ & - \Delta \left[ \mathrm{No.~ of~ AFM~NN~ pairs}\right] \Bigg\}
\end{align}
where $\Delta E_0 \ll \Delta E_\mathrm{mag}$ is assumed based on their similar atomic configurations. Going from the G state to G+E state for the supercell of 32 magnetic Cu atoms as depicted in Fig. \ref{fig:0_0625} involves changing 16 pairs from AFM to FM ordering. Therefore, $\Delta (\mathrm{No.~ of~ FM~ NN~pairs}) = 16$ and $\Delta (\mathrm{No.~ of~ AFM~ NN~pairs}) = -16$, which gives:
\begin{equation}
    \Delta E_\mathrm{G+E,G} (\mathrm{32~Cu~atoms}) = 8J
\end{equation}
We normalize this formula to work with the values listed in Table \ref{tab:energy}:
\begin{equation}
    J = 4\times \left(E_{\mathrm{G+E}}^{\mathrm{per~ atom}} - E_{\mathrm{G}}^{\mathrm{per~ atom}}\right) =  127.08 ~ \mathrm{meV}
\end{equation}
which is comparable with measurements of other cuprates possessing similar superconducting transition temperature ($T_c$) values (Table \ref{tblJ}).

\begin{table}[h]
\caption{The values for magnetic exchange coupling parameter $J$ per Cu atom for cuprate systems.}
    \centering
\begin{ruledtabular}
    \begin{tabular}{l c c r}
      System & $T^\mathrm{max}_c$ (K)  &$J$ (meV) & Remark\\
      \hline
Nd-CCO\footnotemark[1]&  24 & 126, 155 & Expt. \cite{Matsuda1990,Bourges1997} \\
Pr-CCO\footnotemark[1]&  22 & 106, 130, 121 & Expt. \cite{Matsuda1990,Bourges1997,Sumarlin1995} \\
Sm-CCO\footnotemark[1]&  20 & 110 & Expt. \cite{Sulewski} \\
Bi2201\footnotemark[2]& 34 & 143 & Expt. \cite{peng2017influence} \\
La214\footnotemark[2]&  39 &  133 (138) & Expt. \cite{Bourges1997} (DFT \cite{Lane2018}) \\
IL-SLCO\footnotemark[1]&  42 & 127 & DFT (this work) \\
    \end{tabular}
\end{ruledtabular}
\footnotetext[1]{Electron-doped cuprates}
\footnotetext[2]{Hole-doped cuprates}
    \label{tblJ}
\end{table}

\subsubsection{A secondary dopant effect elucidated by combinations of magnetic and nonmagnetic phases at $x=0.1$ and $x=0.125$}

An enhancement of $N(0)$ can be achieved also by intermixing some nonmagnetic planes in the AF system. However, we note that in general, a purely nonmagnetic N state has a higher $N(0)$ than the AF G or C states. There are interesting experimental reports on layered cuprates \cite{Kotegawa2004, Mukuda2006} consisting of a mixture of superconducting and AF-ordered planes. However, our method does not have the capability to precisely distinguish among the nonmagnetic states the configuration that will likely correspond to the superconducting states observed in those reports. Instead, here we will study the appearance of these nonmagnetic Cu sites in our low-energy multiphase systems toward elucidating a secondary role of the La dopant substitution made accessible by our supercell method. 

The substitution of a dopant atom to a parent cuprate system is intended to alter the charge distribution in the CuO$_2$ planes. For example, the replacement of divalent Sr with the trivalent La in SLCO adds an electron to the unit cell, resulting in an electron-doped cuprate system. The additional charge carrier moves into the CuO$_2$ planes and weakens the AF order. If sufficient charge is added, an originally magnetic plane can become a completely nonmagnetic one. Typically, the CuO$_2$ planes closest to the dopant site have their AF order weakened the most, thus resulting in a magnetic moment distribution in which the magnitude increases with distance. This is what we refer as the ``primary" effect of the dopant substitution, seen in experiments involving layered cuprates where equivalency between the so-called ``inner" and ``outer" CuO$_2$ planes are broken in terms of the AF moment size \cite{Kotegawa2004, Mukuda2006}. A gradual distribution of AF strength as a function of distance to the dopant in this fashion is what we also observe in our doped systems in the G or C states. In our low-energy states, however, some deviations from the distribution mentioned in the previous paragraph are noted. 

\begin{figure}[h]
   \centering
		\includegraphics[width=0.8\linewidth]{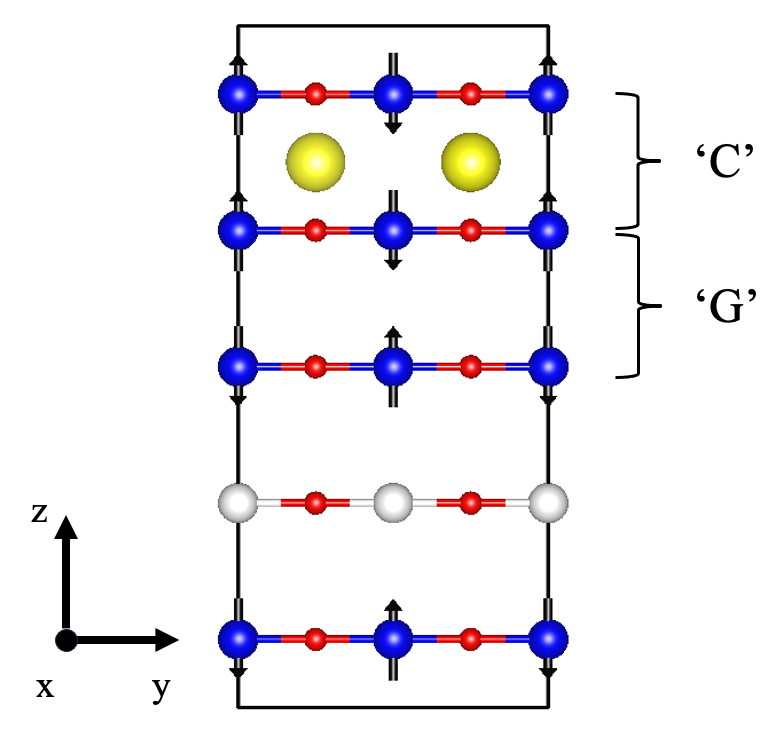}
	\caption{An illustration of the $2\times 2 \times 5$ supercell in the G+C+N state for SLCO with $x=0.1$, viewed along the CuO$_2$ plane. Yellow, red, and blue (white) spheres
represent La, O, and magnetic (nonmagnetic) Cu atoms. Sr atoms are omitted for clarity. The magnetic moments are drawn normal to the CuO$_2$ planes for simplicity.} 
	\label{FIG:GplusCplusN}
\end{figure}

First, we highlight the G+C+N state in a $2\times 2\times5$ supercell in Fig. \ref{FIG:GplusCplusN} located 10.64 meV above the G reference state. It consists of the nearest CuO$_2$ planes surrounding the La dopants in C magnetic configuration ($m\approx \pm 0.33 \mu_B$), sandwiched by planes taking G ordering ($m\approx \pm 0.41 \mu_B$). However, we find a nonmagnetic plane every four CuO$_2$ planes, resulting in a domainlike form instead of a continuous G pattern. The AFM order thus appears inhomogeneously in a short length scale; this may be related to the loss of NMR intensity in Ref. \cite{Williams2005}. $N(0)$ is slightly increased by approximately 22 $\%$ to 0.22 states$/(\mathrm{eV} \cdot \mathrm{spin} \cdot \mathrm{Cu~atom})$ relative to the G state. This G+C+N state is also found when the two La dopants are placed in separate layers in the $\sqrt{2}\times\sqrt{2}\times10$ cell with similar $E$ and $N(0)$ values. Furthermore, we can also compute the the Sommerfeld parameter $\gamma$ of electronic specific heat, defined as \cite{Plakida_2010}:
\begin{equation}
    C_\mathrm{el} (T\rightarrow 0) = \gamma T
\end{equation}
\begin{equation}
    \gamma = \frac{2}{3}\pi^2 k^2_\mathrm{B} N(0)
\end{equation}
where the DOS at Fermi level $N(0)$ is defined per atom and per one spin direction. This quantity in the normal state extrapolated to zero temperature can be estimated from $N(0)$ of a computational cell containing $X$ formula units \cite{Plakida_2010}:
\begin{equation} \label{Sommerfeld}
    \gamma^0_n (\mathrm{mJ}/\mathrm{K}^2 \cdot \mathrm{mol}) \approx 2.36 \times 2N(0) (\mathrm{state}/\mathrm{eV}\cdot X)
\end{equation} 
Ref. \cite{Liu_2005} measured the Sommerfeld parameter of SLCO $(x=0.1)$ to be $1.04 \pm 0.01 (\mathrm{mJ}/\mathrm{K}^2 \cdot \mathrm{mol})$.
We note a reasonable agreement between the experimental value with our estimation of $0.918 (0.869)$ and $0.996 (1.053)$ $\mathrm{mJ}/\mathrm{K}^2 \cdot \mathrm{mol}$ for the G and G+C+N states  with $\sqrt{2}\times\sqrt{2}\times10$ ($2\times2\times5$) cells, respectively. 

\begin{figure}[h]
    \centering
    \includegraphics[width=1\linewidth]{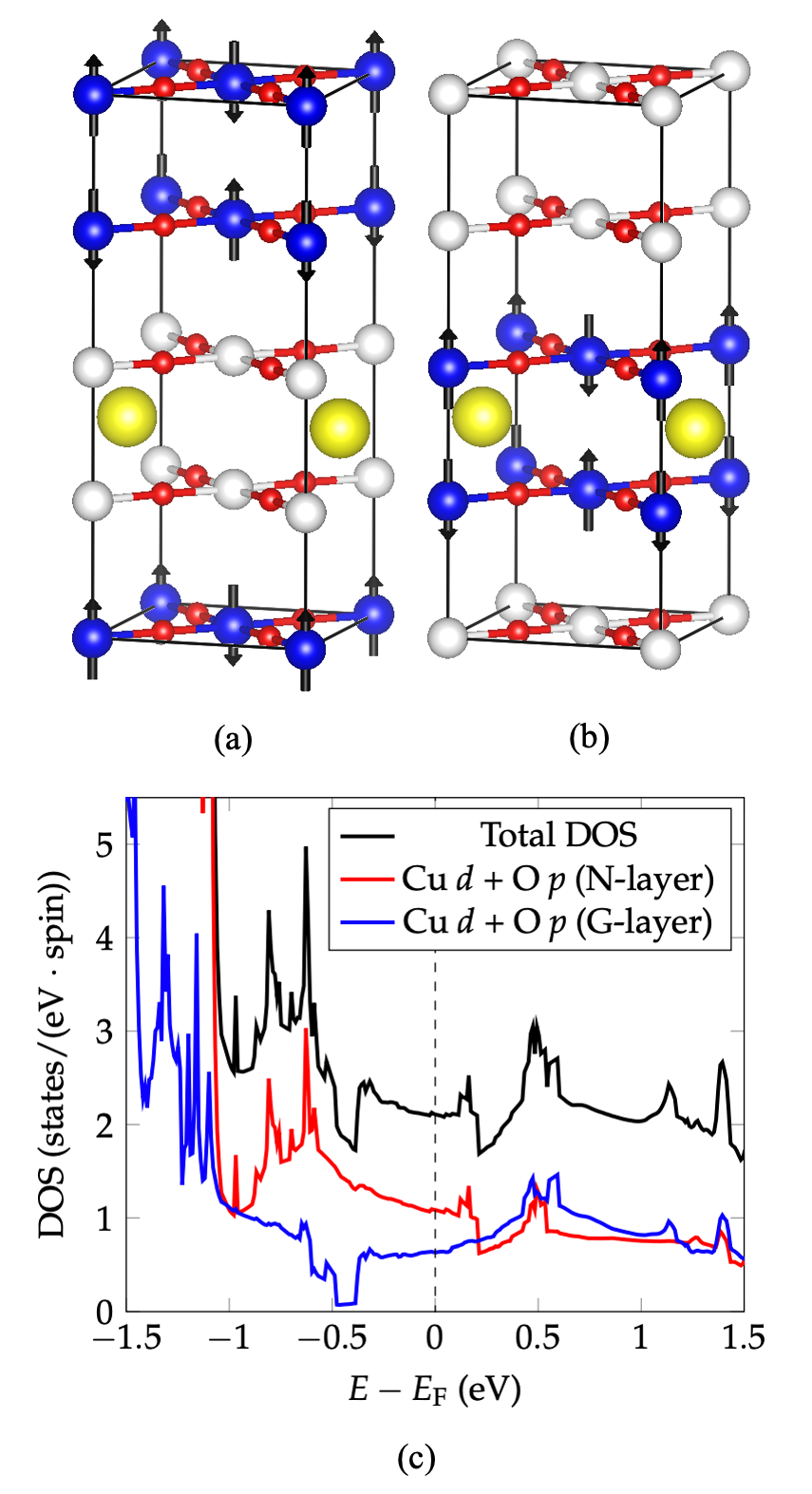}
    \caption{The $\sqrt{2}\times \sqrt{2} \times 4$ supercell in the G+N state for SLCO with $x=0.125$ with the nonmagnetic sites being closer (a) and further (b) from the dopant site. The Sr atoms are omitted for clarity. Yellow, red, and blue (white) spheres represent La, O, and magnetic (nonmagnetic) Cu atoms. The magnetic moments are drawn normal to the CuO$_2$ planes for simplicity. (c): The total DOS (in black) for G+N state from Figure \ref{fig:0_125}(b). The Fermi level is set to zero. Projected DOS to the Cu $d$ and O $p$ orbitals from magnetic and nonmagnetic planes are shown in blue and red traces, respectively. }
    \label{fig:0_125}
\end{figure}

Next, we discuss the G+N state. Figures \ref{fig:0_125}(a) and \ref{fig:0_125}(b) illustrate this state in the $\sqrt{2}\times\sqrt{2}\times4$ cell for $x=0.125$. There are two magnetic planes and two nonmagnetic planes in both cases. However, the configuration with nonmagnetic planes closest to the dopant site [Fig. \ref{fig:0_125}(a)] has a lower total energy than the opposite one depicted in Fig. \ref{fig:0_125}(b) (14.62 meV vs 22.79 meV above the reference state). This may be due to difference in AF strength of the magnetic planes [$m \approx \pm 0.41 \mu_B$ in Fig. \ref{fig:0_125}(a) vs $m \approx \pm 0.33 \mu_B$ in Fig. \ref{fig:0_125}(b)] caused by their respective distance from the dopant site. The DOS for the G+N configuration is shown in Fig. \ref{fig:0_125}(c). Here we take the configuration in  Fig. \ref{fig:0_125}(b) as an example. The nonmagnetic and magnetic layers contribute distinctly to the total DOS. For example, a gap exists around $E-E_\mathrm{F} \approx -0.5 $ eV only on the G layers (blue traces) and not on the N layers (red traces). The gap displacement to a lower energy is consistent with the electron doping condition. We note a $40.6 \% $ increase of $N(0)$ for the G+N state relative to the reference state. The appearance of nonmagnetic planes may reflect the spin dilution effects in which electron doping results in replacement of some Cu$^{2+}$ by Cu$^{+}$ spinless ions, as also suggested in the AMR studies in Ref. \cite{Jovanovic2010}. The location of the nonmagnetic planes away from the dopant site, as depicted in Fig. \ref{fig:0_125}(b), is interesting. It deviates from the ``primary" effect discussed earlier, according to which the closer planes should be more likely to lose its AF order than the further ones. We try to comprehend this in the next paragraph. 

From Fig. \ref{FIG:flowchart}, we may think of achieving the G+N state in Fig. \ref{fig:0_125}(b) starting from the nonmagnetic N state and replacing the nonmagnetic planes closest to the impurity site with magnetic planes. The total energy is thus reduced by the AFM order. To complete this picture, we need to address how magnetism can be introduced near the La impurity site in this case. The substitution of La does not only alter the number of valence electrons. It also breaks the symmetry of the parent compound and introduce disorders in the SLCO structure due to difference in atomic size, bond length, bond angle, etc. with the parent structure. Experimentally, cases of magnetic moments induced by impurities in other cuprates have been reported \cite{Alloul2009}. On the other hand, disorders have also been studied theoretically with $t-J$ model calculations to be capable of pinning local antiferromagnetic regions near impurity sites in underdoped cuprates \cite{Christensen2011}. This subtle, ``secondary" effect of substitutional disorder may not generally manifest itself for structures in the ground state, where the primary effect is expected to be dominant, but it may be observed in higher energy states. This is consistent with our total energy calculations (Table \ref{tab:energy}) for the two G+N states in Fig. \ref{fig:0_125}, where the state in Fig. \ref{fig:0_125}(b) that represents this idea has a slightly higher energy than the state in Fig. \ref{fig:0_125}(a). Nevertheless, we stress that this explanation is only an attempt to understand the results in Fig. \ref{fig:0_125}(b). The task of validating the $t-J$ model applicability of Ref. \cite{Christensen2011} to SLCO system is, however, beyond the scope of this paper. We hope our findings here  would incite further studies of impurity effects in the magnetic phases of cuprates.

Having introduced the low-energy multiphase states in the previous pages, we finish this subsection with a few comments on stability based on total energy calculations in Table \ref{tab:energy}, similar to Ref. \cite{zhang2020}. The highlighted multiphase states in this paper exist in the doped compounds between the lowest energy, single-phase G or C states and the nonmagnetic state. The multiphase states individually have different energies, therefore some are more likely to be converged into (more ``stable") than the others in a variational total energy calculation. We tried comparing them (G+E, G+C+N, etc.) for the same $x$ value to verify this notion. Additional calculations initialized with these states across $x$ values and supercell sizes are thus performed. Although these attempts often converged into the more robust G, C, or N states, we had some success in, for example, obtaining the two types of G+N state in $x=0.1$ with the $2\times 2\times 5$ supercell. Their energies are slightly higher than the G+C+N state as indicated in Table \ref{tab:energy}. In turn, the G+C+N state was also found at $x=0.0625$ with $\sqrt{2}\times\sqrt{2}\times 8$ supercell but with a slightly larger energy difference and no improvement in $N(0)$ to the ground state. The stability of each multiphase state thus varies with $x$, and competition of such states may continue to be an interesting phenomenon for more extensive studies in the future. 

\subsection{\label{SubsecPress}Pressure effects}
Unlike hole-doped cuprates, pressure effects in the electronic structure of electron-doped compounds are reported to be much weaker \cite{murayama1989anomalous,Ishiwata2013, TAKAHASHI1994395}. No pressure dependence on $T_c$ and several other parameters for SLCO are observed up to 1 GPa \cite{DiCastro_2009, KIM2006}. As far as we know, there is only a report of small increase of 3 K ($\approx 7\%$) for SLCO measured at high pressure of 15 GPa in Ref. \cite{Ishiwata2013}. All of the pressure studies involving SLCO were performed with hydrostatic pressure, where both $a$ and $c$ lattice parameters get compressed. Uniaxial pressure studies on this material have not been available.

\begin{figure}[h]
    \centering
    \includegraphics[width=1\linewidth]{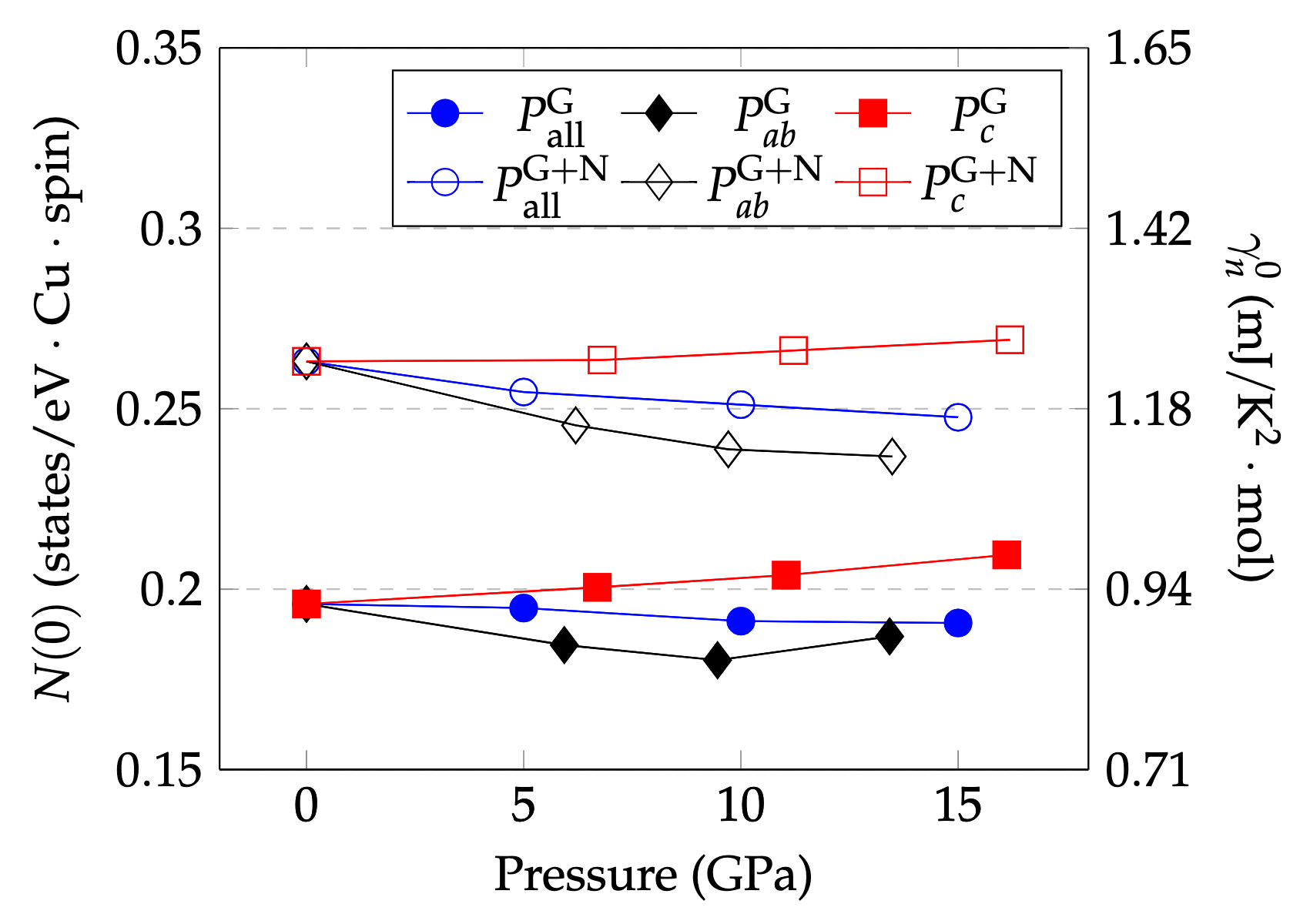}
    \caption{The density of states at Fermi level, $N(0)$ and the estimated specific heat coefficient, $\gamma^0_n$, for Sr$_{0.875}$La$_{0.125}$CuO$_2$ in the G (closed symbols) and G+N (open symbols) states under hydrostatic ($P_\mathrm{all}$), in-plane ($P_{ab}$), and uniaxial $c$-axis ($P_{c}$) compressions.}
    \label{fig:pressure}
\end{figure}

Here, we perform electronic structure calculations for SLCO ($x=0.125$) under hydrostatic ($P_\mathrm{all}$), in-plane ($P_{ab}$), and uniaxial $c$-axis ($P_{c}$) compressions. The former is executed with a built-in feature of the VASP code, while the directional compressions are performed by reducing the respective lattice parameters by hand. The results are illustrated in Fig. \ref{fig:pressure}, where we have plotted the values of $N(0)$ alongside their respective estimation for $\gamma^0_n$ following Eq. \eqref{Sommerfeld} for the G and G+N [Figure \ref{fig:0_125}(b)] states. The pressure effects to $N(0)$ are generally very weak. We note a small increase with uniaxial $c$-axis compression $P_{c}$, but not $P_\mathrm{all}$ or $P_{ab}$. 

Under hydrostatic compressions, the values of $a$ and $c$ are both reduced as shown in Table \ref{tab:hydro} for the G state. The values for G+N state are highly similar. From these results, the effects of in-plane compression to the value of $N(0)$ are likely to be stronger than $c$-axis compression when both are present. We look at these results with the evolution of $a$ and $c$ under doping $x$ shown in Fig. \ref{fig:latticeparam}. Although the La concentration here is fixed at $x=0.125$, the reduced $a$ ($c$) is akin to moving towards lesser (higher) doping. It is possible that introducing La dopant atoms has an effect similar to applying pressure, which will be an interesting issue to be studied in the future. 

\begin{table}[h]
   \caption{The lattice parameters for Sr$_{0.875}$La$_{0.125}$CuO$_2$ under hydrostatic compressions.}
    \label{tab:hydro}
    \centering
\begin{ruledtabular}
    \begin{tabular}{c c c c c}
      Parameter & $0$ & $5$ GPa &$10$ GPa& $15$ GPa \\
      \hline
    $a$ &  3.932 & 3.896  & 3.866  & 3.840 \\
     $c$ &  3.418 & 3.352  & 3.295 & 3.240 \\
    \end{tabular}
\end{ruledtabular}
\end{table}

Although $N(0)$ appears to be slightly increasing with $P_c$, more studies will be necessary to understand the pressure dependence of $T_c$. The changes of $N(0)$ are minute, and the correlation between $N(0)$ and $T_c$ is more subtle in cuprates than in conventional superconductors. On the other hand, the uniaxial pressure dependence for $T_c$ in electron-doped cuprates itself is not yet established. For example, studies of Sm-CCO \cite{KUND1998173} deduced that decreasing $c$ ($a$) has small positive (negative) effects on $T_c$. However, the study for Nd-CCO suggested the opposite effect \cite{KAGA2005442}. The effects of uniaxial compression may also differ in SLCO as it is not isostructural with these compounds. Hence, we encourage further experimental investigations of uniaxial stress in SLCO.

Finally, the electronic states showcased in this brief work are examples which do not form an exhaustive list. It is plausible that other competitive multiphase states may be found with more extensive calculations. Comprehensive studies of these interesting states as precursor candidates may be beneficial for elucidating superconductivity of magnetic origin.

\section{Conclusions}
In this brief work, we study the electron-doped infinite-layer $\mathrm{Sr}_{1-x}\mathrm{La}_{x}\mathrm{CuO}_2$ using first-principles SCAN density functional calculations. The ground-state G-AFM phase in the parent and doped compounds are investigated to confirm the expected doping-induced behaviors known from experiments, such as the metal-insulator transition, lattice parameter evolution, and magnetic anisotropy energies. The main results of this paper arrive from our investigation of low-energy, multiphase states in the doped compounds. Analysis of the G+E state at $x=0.0625$ suggests that magnetic fluctuations may lead to a higher density of states at Fermi level, $N(0)$. Furthermore, our inspection of the multiphase states consisting of magnetic and nonmagnetic CuO$_2$ planes gives an impression that in addition to providing charge carriers that weakens the AF order in the ground state, the La substitution may also induce AF regions in the otherwise nonmagnetic system at higher energies from the disorders that it creates. Our estimation of commonly measured quantities derived from these states, such as the magnetic exchange coupling parameter $J$ and the specific heat coefficient $\gamma$, is in good agreement with experiment. Finally, we note a small positive effect of uniaxial $c$-axis compressions on the value of $N(0)$ for both single-phase and multiphase states. We believe that further studies of exotic magnetic states may help explain the mechanism for various correlated phenomena of these materials in the future.
\begin{acknowledgments}
The calculations were performed with the facilities of the Supercomputer Center, the Institute for Solid State Physics, the University of Tokyo.
\end{acknowledgments}
\bibliography{apssamp}

\providecommand{\noopsort}[1]{}\providecommand{\singleletter}[1]{#1}%
\begin{thebibliography}{56}%
\makeatletter
\providecommand \@ifxundefined [1]{%
 \@ifx{#1\undefined}
}%
\providecommand \@ifnum [1]{%
 \ifnum #1\expandafter \@firstoftwo
 \else \expandafter \@secondoftwo
 \fi
}%
\providecommand \@ifx [1]{%
 \ifx #1\expandafter \@firstoftwo
 \else \expandafter \@secondoftwo
 \fi
}%
\providecommand \natexlab [1]{#1}%
\providecommand \enquote  [1]{``#1''}%
\providecommand \bibnamefont  [1]{#1}%
\providecommand \bibfnamefont [1]{#1}%
\providecommand \citenamefont [1]{#1}%
\providecommand \href@noop [0]{\@secondoftwo}%
\providecommand \href [0]{\begingroup \@sanitize@url \@href}%
\providecommand \@href[1]{\@@startlink{#1}\@@href}%
\providecommand \@@href[1]{\endgroup#1\@@endlink}%
\providecommand \@sanitize@url [0]{\catcode `\\12\catcode `\$12\catcode `\&12\catcode `\#12\catcode `\^12\catcode `\_12\catcode `\%12\relax}%
\providecommand \@@startlink[1]{}%
\providecommand \@@endlink[0]{}%
\providecommand \url  [0]{\begingroup\@sanitize@url \@url }%
\providecommand \@url [1]{\endgroup\@href {#1}{\urlprefix }}%
\providecommand \urlprefix  [0]{URL }%
\providecommand \Eprint [0]{\href }%
\providecommand \doibase [0]{https://doi.org/}%
\providecommand \selectlanguage [0]{\@gobble}%
\providecommand \bibinfo  [0]{\@secondoftwo}%
\providecommand \bibfield  [0]{\@secondoftwo}%
\providecommand \translation [1]{[#1]}%
\providecommand \BibitemOpen [0]{}%
\providecommand \bibitemStop [0]{}%
\providecommand \bibitemNoStop [0]{.\EOS\space}%
\providecommand \EOS [0]{\spacefactor3000\relax}%
\providecommand \BibitemShut  [1]{\csname bibitem#1\endcsname}%
\let\auto@bib@innerbib\@empty
\bibitem [{\citenamefont {Bednorz}\ and\ \citenamefont {M{\"u}ller}(1986)}]{bednorz1986}%
  \BibitemOpen
  \bibfield  {author} {\bibinfo {author} {\bibfnamefont {J.~G.}\ \bibnamefont {Bednorz}}\ and\ \bibinfo {author} {\bibfnamefont {K.~A.}\ \bibnamefont {M{\"u}ller}},\ }\bibfield  {title} {\bibinfo {title} {Possible high $\textrm{T}_c$ superconductivity in the $\textrm{Ba-La-Cu-O}$ system},\ }\href {https://doi.org/https://doi.org/10.1007/BF01303701} {\bibfield  {journal} {\bibinfo  {journal} {Z. Phys. B}\ }\textbf {\bibinfo {volume} {64}},\ \bibinfo {pages} {189} (\bibinfo {year} {1986})}\BibitemShut {NoStop}%
\bibitem [{\citenamefont {Pickett}(1989)}]{Pickett1989}%
  \BibitemOpen
  \bibfield  {author} {\bibinfo {author} {\bibfnamefont {W.~E.}\ \bibnamefont {Pickett}},\ }\bibfield  {title} {\bibinfo {title} {Electronic structure of the high-temperature oxide superconductors},\ }\href {https://doi.org/10.1103/RevModPhys.61.433} {\bibfield  {journal} {\bibinfo  {journal} {Rev. Mod. Phys.}\ }\textbf {\bibinfo {volume} {61}},\ \bibinfo {pages} {433} (\bibinfo {year} {1989})}\BibitemShut {NoStop}%
\bibitem [{\citenamefont {Damascelli}\ \emph {et~al.}(2003)\citenamefont {Damascelli}, \citenamefont {Hussain},\ and\ \citenamefont {Shen}}]{Damascelli2003}%
  \BibitemOpen
  \bibfield  {author} {\bibinfo {author} {\bibfnamefont {A.}~\bibnamefont {Damascelli}}, \bibinfo {author} {\bibfnamefont {Z.}~\bibnamefont {Hussain}},\ and\ \bibinfo {author} {\bibfnamefont {Z.-X.}\ \bibnamefont {Shen}},\ }\bibfield  {title} {\bibinfo {title} {Angle-resolved photoemission studies of the cuprate superconductors},\ }\href {https://doi.org/10.1103/RevModPhys.75.473} {\bibfield  {journal} {\bibinfo  {journal} {Rev. Mod. Phys.}\ }\textbf {\bibinfo {volume} {75}},\ \bibinfo {pages} {473} (\bibinfo {year} {2003})}\BibitemShut {NoStop}%
\bibitem [{\citenamefont {Armitage}\ \emph {et~al.}(2010)\citenamefont {Armitage}, \citenamefont {Fournier},\ and\ \citenamefont {Greene}}]{Armitage2010}%
  \BibitemOpen
  \bibfield  {author} {\bibinfo {author} {\bibfnamefont {N.~P.}\ \bibnamefont {Armitage}}, \bibinfo {author} {\bibfnamefont {P.}~\bibnamefont {Fournier}},\ and\ \bibinfo {author} {\bibfnamefont {R.~L.}\ \bibnamefont {Greene}},\ }\bibfield  {title} {\bibinfo {title} {Progress and perspectives on electron-doped cuprates},\ }\href {https://doi.org/10.1103/RevModPhys.82.2421} {\bibfield  {journal} {\bibinfo  {journal} {Rev. Mod. Phys.}\ }\textbf {\bibinfo {volume} {82}},\ \bibinfo {pages} {2421} (\bibinfo {year} {2010})}\BibitemShut {NoStop}%
\bibitem [{\citenamefont {Fournier}(2015)}]{FOURNIER2015}%
  \BibitemOpen
  \bibfield  {author} {\bibinfo {author} {\bibfnamefont {P.}~\bibnamefont {Fournier}},\ }\bibfield  {title} {\bibinfo {title} {T' and infinite-layer electron-doped cuprates},\ }\href {https://doi.org/https://doi.org/10.1016/j.physc.2015.02.036} {\bibfield  {journal} {\bibinfo  {journal} {Physica C}\ }\textbf {\bibinfo {volume} {514}},\ \bibinfo {pages} {314} (\bibinfo {year} {2015})}\BibitemShut {NoStop}%
\bibitem [{\citenamefont {Naito}\ \emph {et~al.}(2016)\citenamefont {Naito}, \citenamefont {Krockenberger}, \citenamefont {Ikeda},\ and\ \citenamefont {Yamamoto}}]{NAITO2016}%
  \BibitemOpen
  \bibfield  {author} {\bibinfo {author} {\bibfnamefont {M.}~\bibnamefont {Naito}}, \bibinfo {author} {\bibfnamefont {Y.}~\bibnamefont {Krockenberger}}, \bibinfo {author} {\bibfnamefont {A.}~\bibnamefont {Ikeda}},\ and\ \bibinfo {author} {\bibfnamefont {H.}~\bibnamefont {Yamamoto}},\ }\bibfield  {title} {\bibinfo {title} {Reassessment of the electronic state, magnetism, and superconductivity in high-$\mathit{T}_c$ cuprates with the $\mathrm{Nd}_2\mathrm{CuO}_4$ structure},\ }\href {https://doi.org/https://doi.org/10.1016/j.physc.2016.02.012} {\bibfield  {journal} {\bibinfo  {journal} {Physica C}\ }\textbf {\bibinfo {volume} {523}},\ \bibinfo {pages} {28} (\bibinfo {year} {2016})}\BibitemShut {NoStop}%
\bibitem [{\citenamefont {Jovanovi\ifmmode~\acute{c}\else \'{c}\fi{}}\ \emph {et~al.}(2010)\citenamefont {Jovanovi\ifmmode~\acute{c}\else \'{c}\fi{}}, \citenamefont {Fruchter}, \citenamefont {Li},\ and\ \citenamefont {Raffy}}]{Jovanovic2010}%
  \BibitemOpen
  \bibfield  {author} {\bibinfo {author} {\bibfnamefont {V.~P.}\ \bibnamefont {Jovanovi\ifmmode~\acute{c}\else \'{c}\fi{}}}, \bibinfo {author} {\bibfnamefont {L.}~\bibnamefont {Fruchter}}, \bibinfo {author} {\bibfnamefont {Z.~Z.}\ \bibnamefont {Li}},\ and\ \bibinfo {author} {\bibfnamefont {H.}~\bibnamefont {Raffy}},\ }\bibfield  {title} {\bibinfo {title} {Anisotropy of the in-plane angular magnetoresistance of electron-doped $\mathrm{Sr}_{1-x}\mathrm{La}_x\mathrm{CuO}_2$ thin films},\ }\href {https://doi.org/10.1103/PhysRevB.81.134520} {\bibfield  {journal} {\bibinfo  {journal} {Phys. Rev. B}\ }\textbf {\bibinfo {volume} {81}},\ \bibinfo {pages} {134520} (\bibinfo {year} {2010})}\BibitemShut {NoStop}%
\bibitem [{\citenamefont {Jorgensen}\ \emph {et~al.}(1993)\citenamefont {Jorgensen}, \citenamefont {Radaelli}, \citenamefont {Hinks}, \citenamefont {Wagner}, \citenamefont {Kikkawa}, \citenamefont {Er},\ and\ \citenamefont {Kanamaru}}]{Jorgensen1993}%
  \BibitemOpen
  \bibfield  {author} {\bibinfo {author} {\bibfnamefont {J.~D.}\ \bibnamefont {Jorgensen}}, \bibinfo {author} {\bibfnamefont {P.~G.}\ \bibnamefont {Radaelli}}, \bibinfo {author} {\bibfnamefont {D.~G.}\ \bibnamefont {Hinks}}, \bibinfo {author} {\bibfnamefont {J.~L.}\ \bibnamefont {Wagner}}, \bibinfo {author} {\bibfnamefont {S.}~\bibnamefont {Kikkawa}}, \bibinfo {author} {\bibfnamefont {G.}~\bibnamefont {Er}},\ and\ \bibinfo {author} {\bibfnamefont {F.}~\bibnamefont {Kanamaru}},\ }\bibfield  {title} {\bibinfo {title} {Structure of superconducting $\mathrm{Sr}_{0.9}\mathrm{La}_{0.1}\mathrm{CuO}_2$ ($\mathit{T}_c=42 ~\mathrm{K}$) from neutron powder diffraction},\ }\href {https://doi.org/10.1103/PhysRevB.47.14654} {\bibfield  {journal} {\bibinfo  {journal} {Phys. Rev. B}\ }\textbf {\bibinfo {volume} {47}},\ \bibinfo {pages} {14654} (\bibinfo {year} {1993})}\BibitemShut {NoStop}%
\bibitem [{\citenamefont {Jovanovi\ifmmode~\acute{c}\else \'{c}\fi{}}\ \emph {et~al.}(2021)\citenamefont {Jovanovi\ifmmode~\acute{c}\else \'{c}\fi{}}, \citenamefont {Raffy}, \citenamefont {Li}, \citenamefont {Rem\'enyi},\ and\ \citenamefont {Monceau}}]{Jovanovic2021}%
  \BibitemOpen
  \bibfield  {author} {\bibinfo {author} {\bibfnamefont {V.~P.}\ \bibnamefont {Jovanovi\ifmmode~\acute{c}\else \'{c}\fi{}}}, \bibinfo {author} {\bibfnamefont {H.}~\bibnamefont {Raffy}}, \bibinfo {author} {\bibfnamefont {Z.~Z.}\ \bibnamefont {Li}}, \bibinfo {author} {\bibfnamefont {G.}~\bibnamefont {Rem\'enyi}},\ and\ \bibinfo {author} {\bibfnamefont {P.}~\bibnamefont {Monceau}},\ }\bibfield  {title} {\bibinfo {title} {High magnetic-field evolution of the in-plane angular magnetoresistance of electron-doped $\mathrm{Sr}_{1-x}\mathrm{La}_x\mathrm{CuO}_2$ in the normal state},\ }\href {https://doi.org/10.1103/PhysRevB.103.014520} {\bibfield  {journal} {\bibinfo  {journal} {Phys. Rev. B}\ }\textbf {\bibinfo {volume} {103}},\ \bibinfo {pages} {014520} (\bibinfo {year} {2021})}\BibitemShut {NoStop}%
\bibitem [{\citenamefont {Williams}\ \emph {et~al.}(2005)\citenamefont {Williams}, \citenamefont {Haase}, \citenamefont {Park}, \citenamefont {Kim},\ and\ \citenamefont {Lee}}]{Williams2005}%
  \BibitemOpen
  \bibfield  {author} {\bibinfo {author} {\bibfnamefont {G.~V.~M.}\ \bibnamefont {Williams}}, \bibinfo {author} {\bibfnamefont {J.}~\bibnamefont {Haase}}, \bibinfo {author} {\bibfnamefont {M.-S.}\ \bibnamefont {Park}}, \bibinfo {author} {\bibfnamefont {K.~H.}\ \bibnamefont {Kim}},\ and\ \bibinfo {author} {\bibfnamefont {S.-I.}\ \bibnamefont {Lee}},\ }\bibfield  {title} {\bibinfo {title} {Doping-dependent reduction of the $\mathrm{Cu}$ nuclear magnetic resonance intensity in the electron-doped superconductors},\ }\href {https://doi.org/10.1103/PhysRevB.72.212511} {\bibfield  {journal} {\bibinfo  {journal} {Phys. Rev. B}\ }\textbf {\bibinfo {volume} {72}},\ \bibinfo {pages} {212511} (\bibinfo {year} {2005})}\BibitemShut {NoStop}%
\bibitem [{\citenamefont {Satoh}\ \emph {et~al.}(2008)\citenamefont {Satoh}, \citenamefont {Takeshita}, \citenamefont {Koda}, \citenamefont {Kadono}, \citenamefont {Ishida}, \citenamefont {Pyon}, \citenamefont {Sasagawa},\ and\ \citenamefont {Takagi}}]{Satoh2008}%
  \BibitemOpen
  \bibfield  {author} {\bibinfo {author} {\bibfnamefont {K.~H.}\ \bibnamefont {Satoh}}, \bibinfo {author} {\bibfnamefont {S.}~\bibnamefont {Takeshita}}, \bibinfo {author} {\bibfnamefont {A.}~\bibnamefont {Koda}}, \bibinfo {author} {\bibfnamefont {R.}~\bibnamefont {Kadono}}, \bibinfo {author} {\bibfnamefont {K.}~\bibnamefont {Ishida}}, \bibinfo {author} {\bibfnamefont {S.}~\bibnamefont {Pyon}}, \bibinfo {author} {\bibfnamefont {T.}~\bibnamefont {Sasagawa}},\ and\ \bibinfo {author} {\bibfnamefont {H.}~\bibnamefont {Takagi}},\ }\bibfield  {title} {\bibinfo {title} {Fermi-liquid behavior and weakly anisotropic superconductivity in the electron-doped cuprate $\mathrm{Sr}_{1-x}\mathrm{La}_x\mathrm{CuO}_2$},\ }\href {https://doi.org/10.1103/PhysRevB.77.224503} {\bibfield  {journal} {\bibinfo  {journal} {Phys. Rev. B}\ }\textbf {\bibinfo {volume} {77}},\ \bibinfo {pages} {224503} (\bibinfo {year} {2008})}\BibitemShut {NoStop}%
\bibitem [{\citenamefont {Harter}\ \emph {et~al.}(2012)\citenamefont {Harter}, \citenamefont {Maritato}, \citenamefont {Shai}, \citenamefont {Monkman}, \citenamefont {Nie}, \citenamefont {Schlom},\ and\ \citenamefont {Shen}}]{Harter2012}%
  \BibitemOpen
  \bibfield  {author} {\bibinfo {author} {\bibfnamefont {J.~W.}\ \bibnamefont {Harter}}, \bibinfo {author} {\bibfnamefont {L.}~\bibnamefont {Maritato}}, \bibinfo {author} {\bibfnamefont {D.~E.}\ \bibnamefont {Shai}}, \bibinfo {author} {\bibfnamefont {E.~J.}\ \bibnamefont {Monkman}}, \bibinfo {author} {\bibfnamefont {Y.}~\bibnamefont {Nie}}, \bibinfo {author} {\bibfnamefont {D.~G.}\ \bibnamefont {Schlom}},\ and\ \bibinfo {author} {\bibfnamefont {K.~M.}\ \bibnamefont {Shen}},\ }\bibfield  {title} {\bibinfo {title} {Nodeless superconducting phase arising from a strong ($\ensuremath{\pi}$, $\ensuremath{\pi}$) antiferromagnetic phase in the infinite-layer electron-doped $\mathrm{Sr}_{1-x}\mathrm{La}_x\mathrm{CuO}_2$ compound},\ }\href {https://doi.org/10.1103/PhysRevLett.109.267001} {\bibfield  {journal} {\bibinfo  {journal} {Phys. Rev. Lett.}\ }\textbf {\bibinfo {volume} {109}},\ \bibinfo {pages} {267001} (\bibinfo {year} {2012})}\BibitemShut {NoStop}%
\bibitem [{\citenamefont {Harter}\ \emph {et~al.}(2015)\citenamefont {Harter}, \citenamefont {Maritato}, \citenamefont {Shai}, \citenamefont {Monkman}, \citenamefont {Nie}, \citenamefont {Schlom},\ and\ \citenamefont {Shen}}]{Harter2015}%
  \BibitemOpen
  \bibfield  {author} {\bibinfo {author} {\bibfnamefont {J.~W.}\ \bibnamefont {Harter}}, \bibinfo {author} {\bibfnamefont {L.}~\bibnamefont {Maritato}}, \bibinfo {author} {\bibfnamefont {D.~E.}\ \bibnamefont {Shai}}, \bibinfo {author} {\bibfnamefont {E.~J.}\ \bibnamefont {Monkman}}, \bibinfo {author} {\bibfnamefont {Y.}~\bibnamefont {Nie}}, \bibinfo {author} {\bibfnamefont {D.~G.}\ \bibnamefont {Schlom}},\ and\ \bibinfo {author} {\bibfnamefont {K.~M.}\ \bibnamefont {Shen}},\ }\bibfield  {title} {\bibinfo {title} {Doping evolution and polar surface reconstruction of the infinite-layer cuprate $\mathrm{Sr}_{1-x}\mathrm{La}_x\mathrm{CuO}_2$},\ }\href {https://doi.org/10.1103/PhysRevB.92.035149} {\bibfield  {journal} {\bibinfo  {journal} {Phys. Rev. B}\ }\textbf {\bibinfo {volume} {92}},\ \bibinfo {pages} {035149} (\bibinfo {year} {2015})}\BibitemShut {NoStop}%
\bibitem [{\citenamefont {Yu}\ \emph {et~al.}(2007)\citenamefont {Yu}, \citenamefont {Higgins}, \citenamefont {Bach},\ and\ \citenamefont {Greene}}]{Yu2007}%
  \BibitemOpen
  \bibfield  {author} {\bibinfo {author} {\bibfnamefont {W.}~\bibnamefont {Yu}}, \bibinfo {author} {\bibfnamefont {J.~S.}\ \bibnamefont {Higgins}}, \bibinfo {author} {\bibfnamefont {P.}~\bibnamefont {Bach}},\ and\ \bibinfo {author} {\bibfnamefont {R.~L.}\ \bibnamefont {Greene}},\ }\bibfield  {title} {\bibinfo {title} {Transport evidence of a magnetic quantum phase transition in electron-doped high-temperature superconductors},\ }\href {https://doi.org/10.1103/PhysRevB.76.020503} {\bibfield  {journal} {\bibinfo  {journal} {Phys. Rev. B}\ }\textbf {\bibinfo {volume} {76}},\ \bibinfo {pages} {020503(R)} (\bibinfo {year} {2007})}\BibitemShut {NoStop}%
\bibitem [{\citenamefont {Sun}\ \emph {et~al.}(2015)\citenamefont {Sun}, \citenamefont {Ruzsinszky},\ and\ \citenamefont {Perdew}}]{Sun2015}%
  \BibitemOpen
  \bibfield  {author} {\bibinfo {author} {\bibfnamefont {J.}~\bibnamefont {Sun}}, \bibinfo {author} {\bibfnamefont {A.}~\bibnamefont {Ruzsinszky}},\ and\ \bibinfo {author} {\bibfnamefont {J.~P.}\ \bibnamefont {Perdew}},\ }\bibfield  {title} {\bibinfo {title} {Strongly constrained and appropriately normed semilocal density functional},\ }\href {https://doi.org/10.1103/PhysRevLett.115.036402} {\bibfield  {journal} {\bibinfo  {journal} {Phys. Rev. Lett.}\ }\textbf {\bibinfo {volume} {115}},\ \bibinfo {pages} {036402} (\bibinfo {year} {2015})}\BibitemShut {NoStop}%
\bibitem [{\citenamefont {Furness}\ \emph {et~al.}(2018)\citenamefont {Furness}, \citenamefont {Zhang}, \citenamefont {Lane}, \citenamefont {Buda}, \citenamefont {Barbiellini}, \citenamefont {Markiewicz}, \citenamefont {Bansil},\ and\ \citenamefont {Sun}}]{furness2018}%
  \BibitemOpen
  \bibfield  {author} {\bibinfo {author} {\bibfnamefont {J.~W.}\ \bibnamefont {Furness}}, \bibinfo {author} {\bibfnamefont {Y.}~\bibnamefont {Zhang}}, \bibinfo {author} {\bibfnamefont {C.}~\bibnamefont {Lane}}, \bibinfo {author} {\bibfnamefont {I.~G.}\ \bibnamefont {Buda}}, \bibinfo {author} {\bibfnamefont {B.}~\bibnamefont {Barbiellini}}, \bibinfo {author} {\bibfnamefont {R.~S.}\ \bibnamefont {Markiewicz}}, \bibinfo {author} {\bibfnamefont {A.}~\bibnamefont {Bansil}},\ and\ \bibinfo {author} {\bibfnamefont {J.}~\bibnamefont {Sun}},\ }\bibfield  {title} {\bibinfo {title} {An accurate first-principles treatment of doping-dependent electronic structure of high-temperature cuprate superconductors},\ }\href {https://doi.org/https://doi.org/10.1038/s42005-018-0009-4} {\bibfield  {journal} {\bibinfo  {journal} {Commun. Phys.}\ }\textbf {\bibinfo {volume} {1}},\ \bibinfo {pages} {11} (\bibinfo {year} {2018})}\BibitemShut {NoStop}%
\bibitem [{\citenamefont {Lane}\ \emph {et~al.}(2018)\citenamefont {Lane}, \citenamefont {Furness}, \citenamefont {Buda}, \citenamefont {Zhang}, \citenamefont {Markiewicz}, \citenamefont {Barbiellini}, \citenamefont {Sun},\ and\ \citenamefont {Bansil}}]{Lane2018}%
  \BibitemOpen
  \bibfield  {author} {\bibinfo {author} {\bibfnamefont {C.}~\bibnamefont {Lane}}, \bibinfo {author} {\bibfnamefont {J.~W.}\ \bibnamefont {Furness}}, \bibinfo {author} {\bibfnamefont {I.~G.}\ \bibnamefont {Buda}}, \bibinfo {author} {\bibfnamefont {Y.}~\bibnamefont {Zhang}}, \bibinfo {author} {\bibfnamefont {R.~S.}\ \bibnamefont {Markiewicz}}, \bibinfo {author} {\bibfnamefont {B.}~\bibnamefont {Barbiellini}}, \bibinfo {author} {\bibfnamefont {J.}~\bibnamefont {Sun}},\ and\ \bibinfo {author} {\bibfnamefont {A.}~\bibnamefont {Bansil}},\ }\bibfield  {title} {\bibinfo {title} {Antiferromagnetic ground state of $\mathrm{La}_{2}\mathrm{CuO}_{4}$: A parameter-free ab initio description},\ }\href {https://doi.org/10.1103/PhysRevB.98.125140} {\bibfield  {journal} {\bibinfo  {journal} {Phys. Rev. B}\ }\textbf {\bibinfo {volume} {98}},\ \bibinfo {pages} {125140} (\bibinfo {year} {2018})}\BibitemShut {NoStop}%
\bibitem [{\citenamefont {Zhang}\ \emph {et~al.}(2020)\citenamefont {Zhang}, \citenamefont {Lane}, \citenamefont {Furness}, \citenamefont {Barbiellini}, \citenamefont {Perdew}, \citenamefont {Markiewicz}, \citenamefont {Bansil},\ and\ \citenamefont {Sun}}]{zhang2020}%
  \BibitemOpen
  \bibfield  {author} {\bibinfo {author} {\bibfnamefont {Y.}~\bibnamefont {Zhang}}, \bibinfo {author} {\bibfnamefont {C.}~\bibnamefont {Lane}}, \bibinfo {author} {\bibfnamefont {J.~W.}\ \bibnamefont {Furness}}, \bibinfo {author} {\bibfnamefont {B.}~\bibnamefont {Barbiellini}}, \bibinfo {author} {\bibfnamefont {J.~P.}\ \bibnamefont {Perdew}}, \bibinfo {author} {\bibfnamefont {R.~S.}\ \bibnamefont {Markiewicz}}, \bibinfo {author} {\bibfnamefont {A.}~\bibnamefont {Bansil}},\ and\ \bibinfo {author} {\bibfnamefont {J.}~\bibnamefont {Sun}},\ }\bibfield  {title} {\bibinfo {title} {Competing stripe and magnetic phases in the cuprates from first principles},\ }\href {https://doi.org/10.1073/pnas.1910411116} {\bibfield  {journal} {\bibinfo  {journal} {Proc. of the Nat'l Acad. of Sci. USA}\ }\textbf {\bibinfo {volume} {117}},\ \bibinfo {pages} {68} (\bibinfo {year} {2020})}\BibitemShut {NoStop}%
\bibitem [{\citenamefont {Nokelainen}\ \emph {et~al.}(2020)\citenamefont {Nokelainen}, \citenamefont {Lane}, \citenamefont {Markiewicz}, \citenamefont {Barbiellini}, \citenamefont {Pulkkinen}, \citenamefont {Singh}, \citenamefont {Sun}, \citenamefont {Pussi},\ and\ \citenamefont {Bansil}}]{Nokelainen2020}%
  \BibitemOpen
  \bibfield  {author} {\bibinfo {author} {\bibfnamefont {J.}~\bibnamefont {Nokelainen}}, \bibinfo {author} {\bibfnamefont {C.}~\bibnamefont {Lane}}, \bibinfo {author} {\bibfnamefont {R.~S.}\ \bibnamefont {Markiewicz}}, \bibinfo {author} {\bibfnamefont {B.}~\bibnamefont {Barbiellini}}, \bibinfo {author} {\bibfnamefont {A.}~\bibnamefont {Pulkkinen}}, \bibinfo {author} {\bibfnamefont {B.}~\bibnamefont {Singh}}, \bibinfo {author} {\bibfnamefont {J.}~\bibnamefont {Sun}}, \bibinfo {author} {\bibfnamefont {K.}~\bibnamefont {Pussi}},\ and\ \bibinfo {author} {\bibfnamefont {A.}~\bibnamefont {Bansil}},\ }\bibfield  {title} {\bibinfo {title} {Ab initio description of the $\mathrm{Bi}_{2}\mathrm{Sr}_{2}\mathrm{CaCu}_{2}\mathrm{O}_{8+\delta}$ electronic structure},\ }\href {https://doi.org/10.1103/PhysRevB.101.214523} {\bibfield  {journal} {\bibinfo  {journal} {Phys. Rev. B}\ }\textbf {\bibinfo {volume} {101}},\ \bibinfo {pages} {214523} (\bibinfo {year} {2020})}\BibitemShut {NoStop}%
\bibitem [{\citenamefont {Tatan}\ \emph {et~al.}(2022)\citenamefont {Tatan}, \citenamefont {Haruyama},\ and\ \citenamefont {Sugino}}]{Tatan2022}%
  \BibitemOpen
  \bibfield  {author} {\bibinfo {author} {\bibfnamefont {A.~N.}\ \bibnamefont {Tatan}}, \bibinfo {author} {\bibfnamefont {J.}~\bibnamefont {Haruyama}},\ and\ \bibinfo {author} {\bibfnamefont {O.}~\bibnamefont {Sugino}},\ }\bibfield  {title} {\bibinfo {title} {{First-principles electronic structure investigation of $\mathrm{HgBa}_2\mathrm{Ca}_{n-1}\mathrm{Cu}_n\mathrm{O}_{2n+2+x}$ with the SCAN density functional}},\ }\href {https://doi.org/10.1063/5.0098554} {\bibfield  {journal} {\bibinfo  {journal} {AIP Advances}\ }\textbf {\bibinfo {volume} {12}},\ \bibinfo {pages} {105308} (\bibinfo {year} {2022})}\BibitemShut {NoStop}%
\bibitem [{\citenamefont {Pokharel}\ \emph {et~al.}(2022)\citenamefont {Pokharel}, \citenamefont {Lane}, \citenamefont {Furness}, \citenamefont {Zhang}, \citenamefont {Ning}, \citenamefont {Barbiellini}, \citenamefont {Markiewicz}, \citenamefont {Zhang}, \citenamefont {Bansil},\ and\ \citenamefont {Sun}}]{pokharel2022sensitivity}%
  \BibitemOpen
  \bibfield  {author} {\bibinfo {author} {\bibfnamefont {K.}~\bibnamefont {Pokharel}}, \bibinfo {author} {\bibfnamefont {C.}~\bibnamefont {Lane}}, \bibinfo {author} {\bibfnamefont {J.~W.}\ \bibnamefont {Furness}}, \bibinfo {author} {\bibfnamefont {R.}~\bibnamefont {Zhang}}, \bibinfo {author} {\bibfnamefont {J.}~\bibnamefont {Ning}}, \bibinfo {author} {\bibfnamefont {B.}~\bibnamefont {Barbiellini}}, \bibinfo {author} {\bibfnamefont {R.~S.}\ \bibnamefont {Markiewicz}}, \bibinfo {author} {\bibfnamefont {Y.}~\bibnamefont {Zhang}}, \bibinfo {author} {\bibfnamefont {A.}~\bibnamefont {Bansil}},\ and\ \bibinfo {author} {\bibfnamefont {J.}~\bibnamefont {Sun}},\ }\bibfield  {title} {\bibinfo {title} {Sensitivity of the electronic and magnetic structures of cuprate superconductors to density functional approximations},\ }\href {https://doi.org/https://doi.org/10.1038/s41524-022-00711-z} {\bibfield  {journal} {\bibinfo  {journal} {npj Comput. Mater.}\ }\textbf {\bibinfo {volume} {8}},\ \bibinfo {pages} {31} (\bibinfo
  {year} {2022})}\BibitemShut {NoStop}%
\bibitem [{\citenamefont {Krukau}\ \emph {et~al.}(2006)\citenamefont {Krukau}, \citenamefont {Vydrov}, \citenamefont {Izmaylov},\ and\ \citenamefont {Scuseria}}]{HSE06}%
  \BibitemOpen
  \bibfield  {author} {\bibinfo {author} {\bibfnamefont {A.~V.}\ \bibnamefont {Krukau}}, \bibinfo {author} {\bibfnamefont {O.~A.}\ \bibnamefont {Vydrov}}, \bibinfo {author} {\bibfnamefont {A.~F.}\ \bibnamefont {Izmaylov}},\ and\ \bibinfo {author} {\bibfnamefont {G.~E.}\ \bibnamefont {Scuseria}},\ }\bibfield  {title} {\bibinfo {title} {{Influence of the exchange screening parameter on the performance of screened hybrid functionals}},\ }\href {https://doi.org/10.1063/1.2404663} {\bibfield  {journal} {\bibinfo  {journal} {The Journal of Chemical Physics}\ }\textbf {\bibinfo {volume} {125}},\ \bibinfo {pages} {224106} (\bibinfo {year} {2006})}\BibitemShut {NoStop}%
\bibitem [{\citenamefont {Deng}\ \emph {et~al.}(2019)\citenamefont {Deng}, \citenamefont {Zheng}, \citenamefont {Wu}, \citenamefont {Huyan}, \citenamefont {Wu}, \citenamefont {Nie}, \citenamefont {Cho},\ and\ \citenamefont {Chu}}]{Deng2019}%
  \BibitemOpen
  \bibfield  {author} {\bibinfo {author} {\bibfnamefont {L.}~\bibnamefont {Deng}}, \bibinfo {author} {\bibfnamefont {Y.}~\bibnamefont {Zheng}}, \bibinfo {author} {\bibfnamefont {Z.}~\bibnamefont {Wu}}, \bibinfo {author} {\bibfnamefont {S.}~\bibnamefont {Huyan}}, \bibinfo {author} {\bibfnamefont {H.-C.}\ \bibnamefont {Wu}}, \bibinfo {author} {\bibfnamefont {Y.}~\bibnamefont {Nie}}, \bibinfo {author} {\bibfnamefont {K.}~\bibnamefont {Cho}},\ and\ \bibinfo {author} {\bibfnamefont {C.-W.}\ \bibnamefont {Chu}},\ }\bibfield  {title} {\bibinfo {title} {Higher superconducting transition temperature by breaking the universal pressure relation},\ }\href {https://doi.org/10.1073/pnas.1819512116} {\bibfield  {journal} {\bibinfo  {journal} {Proc. of the Nat'l Acad. of Sci. USA}\ }\textbf {\bibinfo {volume} {116}},\ \bibinfo {pages} {2004} (\bibinfo {year} {2019})}\BibitemShut {NoStop}%
\bibitem [{\citenamefont {Sboychakov}\ \emph {et~al.}(2008)\citenamefont {Sboychakov}, \citenamefont {Savel'ev}, \citenamefont {Rakhmanov}, \citenamefont {Kugel},\ and\ \citenamefont {Nori}}]{Sboychakov2008}%
  \BibitemOpen
  \bibfield  {author} {\bibinfo {author} {\bibfnamefont {A.~O.}\ \bibnamefont {Sboychakov}}, \bibinfo {author} {\bibfnamefont {S.}~\bibnamefont {Savel'ev}}, \bibinfo {author} {\bibfnamefont {A.~L.}\ \bibnamefont {Rakhmanov}}, \bibinfo {author} {\bibfnamefont {K.~I.}\ \bibnamefont {Kugel}},\ and\ \bibinfo {author} {\bibfnamefont {F.}~\bibnamefont {Nori}},\ }\bibfield  {title} {\bibinfo {title} {Mechanism for phase separation in cuprates and related multiband systems},\ }\href {https://doi.org/10.1103/PhysRevB.77.224504} {\bibfield  {journal} {\bibinfo  {journal} {Phys. Rev. B}\ }\textbf {\bibinfo {volume} {77}},\ \bibinfo {pages} {224504} (\bibinfo {year} {2008})}\BibitemShut {NoStop}%
\bibitem [{\citenamefont {Dagotto}\ \emph {et~al.}(2003)\citenamefont {Dagotto}, \citenamefont {Burgy},\ and\ \citenamefont {Moreo}}]{Dagotto2003}%
  \BibitemOpen
  \bibfield  {author} {\bibinfo {author} {\bibfnamefont {E.}~\bibnamefont {Dagotto}}, \bibinfo {author} {\bibfnamefont {J.}~\bibnamefont {Burgy}},\ and\ \bibinfo {author} {\bibfnamefont {A.}~\bibnamefont {Moreo}},\ }\bibfield  {title} {\bibinfo {title} {Nanoscale phase separation in colossal magnetoresistance materials: lessons for the cuprates?},\ }\href {https://doi.org/https://doi.org/10.1016/S0038-1098(02)00662-2} {\bibfield  {journal} {\bibinfo  {journal} {Solid State Commun.}\ }\textbf {\bibinfo {volume} {126}},\ \bibinfo {pages} {9} (\bibinfo {year} {2003})}\BibitemShut {NoStop}%
\bibitem [{\citenamefont {Kresse}\ and\ \citenamefont {Joubert}(1999)}]{Kresse1999}%
  \BibitemOpen
  \bibfield  {author} {\bibinfo {author} {\bibfnamefont {G.}~\bibnamefont {Kresse}}\ and\ \bibinfo {author} {\bibfnamefont {D.}~\bibnamefont {Joubert}},\ }\bibfield  {title} {\bibinfo {title} {From ultrasoft pseudopotentials to the projector augmented-wave method},\ }\href {https://doi.org/10.1103/PhysRevB.59.1758} {\bibfield  {journal} {\bibinfo  {journal} {Phys. Rev. B}\ }\textbf {\bibinfo {volume} {59}},\ \bibinfo {pages} {1758} (\bibinfo {year} {1999})}\BibitemShut {NoStop}%
\bibitem [{\citenamefont {Kresse}\ and\ \citenamefont {Hafner}(1993)}]{Kresse1993}%
  \BibitemOpen
  \bibfield  {author} {\bibinfo {author} {\bibfnamefont {G.}~\bibnamefont {Kresse}}\ and\ \bibinfo {author} {\bibfnamefont {J.}~\bibnamefont {Hafner}},\ }\bibfield  {title} {\bibinfo {title} {Ab initio molecular dynamics for open-shell transition metals},\ }\href {https://doi.org/10.1103/PhysRevB.48.13115} {\bibfield  {journal} {\bibinfo  {journal} {Phys. Rev. B}\ }\textbf {\bibinfo {volume} {48}},\ \bibinfo {pages} {13115} (\bibinfo {year} {1993})}\BibitemShut {NoStop}%
\bibitem [{\citenamefont {Kresse}\ and\ \citenamefont {Furthm\"uller}(1996)}]{Kresse1996}%
  \BibitemOpen
  \bibfield  {author} {\bibinfo {author} {\bibfnamefont {G.}~\bibnamefont {Kresse}}\ and\ \bibinfo {author} {\bibfnamefont {J.}~\bibnamefont {Furthm\"uller}},\ }\bibfield  {title} {\bibinfo {title} {Efficient iterative schemes for ab initio total-energy calculations using a plane-wave basis set},\ }\href {https://doi.org/10.1103/PhysRevB.54.11169} {\bibfield  {journal} {\bibinfo  {journal} {Phys. Rev. B}\ }\textbf {\bibinfo {volume} {54}},\ \bibinfo {pages} {11169} (\bibinfo {year} {1996})}\BibitemShut {NoStop}%
\bibitem [{\citenamefont {Bl\"ochl}\ \emph {et~al.}(1994)\citenamefont {Bl\"ochl}, \citenamefont {Jepsen},\ and\ \citenamefont {Andersen}}]{Blochl1994}%
  \BibitemOpen
  \bibfield  {author} {\bibinfo {author} {\bibfnamefont {P.~E.}\ \bibnamefont {Bl\"ochl}}, \bibinfo {author} {\bibfnamefont {O.}~\bibnamefont {Jepsen}},\ and\ \bibinfo {author} {\bibfnamefont {O.~K.}\ \bibnamefont {Andersen}},\ }\bibfield  {title} {\bibinfo {title} {Improved tetrahedron method for brillouin-zone integrations},\ }\href {https://doi.org/10.1103/PhysRevB.49.16223} {\bibfield  {journal} {\bibinfo  {journal} {Phys. Rev. B}\ }\textbf {\bibinfo {volume} {49}},\ \bibinfo {pages} {16223} (\bibinfo {year} {1994})}\BibitemShut {NoStop}%
\bibitem [{\citenamefont {Wollan}\ and\ \citenamefont {Koehler}(1955)}]{Wollan1955}%
  \BibitemOpen
  \bibfield  {author} {\bibinfo {author} {\bibfnamefont {E.~O.}\ \bibnamefont {Wollan}}\ and\ \bibinfo {author} {\bibfnamefont {W.~C.}\ \bibnamefont {Koehler}},\ }\bibfield  {title} {\bibinfo {title} {Neutron diffraction study of the magnetic properties of the series of perovskite-type compounds $[(1-x)\mathrm{La}, x\mathrm{Ca}]\mathrm{Mn}\mathrm{O}_{3}$},\ }\href {https://doi.org/10.1103/PhysRev.100.545} {\bibfield  {journal} {\bibinfo  {journal} {Phys. Rev.}\ }\textbf {\bibinfo {volume} {100}},\ \bibinfo {pages} {545} (\bibinfo {year} {1955})}\BibitemShut {NoStop}%
\bibitem [{\citenamefont {Zhang}\ \emph {et~al.}(2021)\citenamefont {Zhang}, \citenamefont {Lane}, \citenamefont {Singh}, \citenamefont {Nokelainen}, \citenamefont {Barbiellini}, \citenamefont {Markiewicz}, \citenamefont {Bansil},\ and\ \citenamefont {Sun}}]{zhang2021nickelate}%
  \BibitemOpen
  \bibfield  {author} {\bibinfo {author} {\bibfnamefont {R.}~\bibnamefont {Zhang}}, \bibinfo {author} {\bibfnamefont {C.}~\bibnamefont {Lane}}, \bibinfo {author} {\bibfnamefont {B.}~\bibnamefont {Singh}}, \bibinfo {author} {\bibfnamefont {J.}~\bibnamefont {Nokelainen}}, \bibinfo {author} {\bibfnamefont {B.}~\bibnamefont {Barbiellini}}, \bibinfo {author} {\bibfnamefont {R.~S.}\ \bibnamefont {Markiewicz}}, \bibinfo {author} {\bibfnamefont {A.}~\bibnamefont {Bansil}},\ and\ \bibinfo {author} {\bibfnamefont {J.}~\bibnamefont {Sun}},\ }\bibfield  {title} {\bibinfo {title} {Magnetic and f-electron effects in $\mathrm{LaNiO}_2$ and $\mathrm{NdNiO}_2$ nickelates with cuprate-like $3d_{x^2-y^2}$ band},\ }\href {https://doi.org/https://doi.org/10.1038/s42005-021-00621-4} {\bibfield  {journal} {\bibinfo  {journal} {Commun. Phys.}\ }\textbf {\bibinfo {volume} {4}},\ \bibinfo {pages} {118} (\bibinfo {year} {2021})}\BibitemShut {NoStop}%
\bibitem [{\citenamefont {Ma}\ and\ \citenamefont {Dudarev}(2015)}]{Ma_constrained}%
  \BibitemOpen
  \bibfield  {author} {\bibinfo {author} {\bibfnamefont {P.-W.}\ \bibnamefont {Ma}}\ and\ \bibinfo {author} {\bibfnamefont {S.~L.}\ \bibnamefont {Dudarev}},\ }\bibfield  {title} {\bibinfo {title} {Constrained density functional for noncollinear magnetism},\ }\href {https://doi.org/10.1103/PhysRevB.91.054420} {\bibfield  {journal} {\bibinfo  {journal} {Phys. Rev. B}\ }\textbf {\bibinfo {volume} {91}},\ \bibinfo {pages} {054420} (\bibinfo {year} {2015})}\BibitemShut {NoStop}%
\bibitem [{\citenamefont {Dagotto}(2003)}]{dagotto2003book}%
  \BibitemOpen
  \bibfield  {author} {\bibinfo {author} {\bibfnamefont {E.}~\bibnamefont {Dagotto}},\ }\href {https://doi.org/https://doi.org/10.1007/978-3-662-05244-0} {\emph {\bibinfo {title} {Nanoscale phase separation and colossal magnetoresistance: the physics of manganites and related compounds}}}\ (\bibinfo  {publisher} {Springer Science \& Business Media},\ \bibinfo {year} {2003})\BibitemShut {NoStop}%
\bibitem [{\citenamefont {Er}\ \emph {et~al.}(1992)\citenamefont {Er}, \citenamefont {Kikkawa}, \citenamefont {Kanamaru}, \citenamefont {Miyamoto}, \citenamefont {Tanaka}, \citenamefont {Sera}, \citenamefont {Sato}, \citenamefont {Hiroi}, \citenamefont {Takano},\ and\ \citenamefont {Bando}}]{kikkawa1992}%
  \BibitemOpen
  \bibfield  {author} {\bibinfo {author} {\bibfnamefont {G.}~\bibnamefont {Er}}, \bibinfo {author} {\bibfnamefont {S.}~\bibnamefont {Kikkawa}}, \bibinfo {author} {\bibfnamefont {F.}~\bibnamefont {Kanamaru}}, \bibinfo {author} {\bibfnamefont {Y.}~\bibnamefont {Miyamoto}}, \bibinfo {author} {\bibfnamefont {S.}~\bibnamefont {Tanaka}}, \bibinfo {author} {\bibfnamefont {M.}~\bibnamefont {Sera}}, \bibinfo {author} {\bibfnamefont {M.}~\bibnamefont {Sato}}, \bibinfo {author} {\bibfnamefont {Z.}~\bibnamefont {Hiroi}}, \bibinfo {author} {\bibfnamefont {M.}~\bibnamefont {Takano}},\ and\ \bibinfo {author} {\bibfnamefont {Y.}~\bibnamefont {Bando}},\ }\bibfield  {title} {\bibinfo {title} {Structural, electrical and magnetic studies of infinite-layered $\mathrm{Sr}_{1-x}\mathrm{La}_x\mathrm{CuO}_2$ superconductor},\ }\href {https://doi.org/https://doi.org/10.1016/0921-4534(92)90446-J} {\bibfield  {journal} {\bibinfo  {journal} {Physica C}\ }\textbf {\bibinfo {volume} {196}},\ \bibinfo {pages} {271} (\bibinfo {year}
  {1992})}\BibitemShut {NoStop}%
\bibitem [{\citenamefont {Ganose}\ \emph {et~al.}(2018)\citenamefont {Ganose}, \citenamefont {Jackson},\ and\ \citenamefont {Scanlon}}]{Ganose2018}%
  \BibitemOpen
  \bibfield  {author} {\bibinfo {author} {\bibfnamefont {A.~M.}\ \bibnamefont {Ganose}}, \bibinfo {author} {\bibfnamefont {A.~J.}\ \bibnamefont {Jackson}},\ and\ \bibinfo {author} {\bibfnamefont {D.~O.}\ \bibnamefont {Scanlon}},\ }\bibfield  {title} {\bibinfo {title} {sumo: Command-line tools for plotting and analysis of periodic \textit{ab initio} calculations},\ }\href {https://doi.org/10.21105/joss.00717} {\bibfield  {journal} {\bibinfo  {journal} {Journ. of Open Source Software}\ }\textbf {\bibinfo {volume} {3}},\ \bibinfo {pages} {717} (\bibinfo {year} {2018})}\BibitemShut {NoStop}%
\bibitem [{\citenamefont {Er}\ \emph {et~al.}(1991)\citenamefont {Er}, \citenamefont {Miyamoto}, \citenamefont {Kanamaru},\ and\ \citenamefont {Kikkawa}}]{ER1991206}%
  \BibitemOpen
  \bibfield  {author} {\bibinfo {author} {\bibfnamefont {G.}~\bibnamefont {Er}}, \bibinfo {author} {\bibfnamefont {Y.}~\bibnamefont {Miyamoto}}, \bibinfo {author} {\bibfnamefont {F.}~\bibnamefont {Kanamaru}},\ and\ \bibinfo {author} {\bibfnamefont {S.}~\bibnamefont {Kikkawa}},\ }\bibfield  {title} {\bibinfo {title} {Superconductivity in the infinite-layer compound $\mathrm{Sr}_{1-x}\mathrm{La}_x\mathrm{CuO}_2$ prepared under high pressure},\ }\href {https://doi.org/https://doi.org/10.1016/0921-4534(91)90356-4} {\bibfield  {journal} {\bibinfo  {journal} {Physica C}\ }\textbf {\bibinfo {volume} {181}},\ \bibinfo {pages} {206} (\bibinfo {year} {1991})}\BibitemShut {NoStop}%
\bibitem [{\citenamefont {Boehm}\ \emph {et~al.}(1998)\citenamefont {Boehm}, \citenamefont {Coad}, \citenamefont {Roessli}, \citenamefont {Zheludev}, \citenamefont {Zolliker}, \citenamefont {Böni}, \citenamefont {Paul}, \citenamefont {Eisaki}, \citenamefont {Motoyama},\ and\ \citenamefont {Uchida}}]{Boehm_1998}%
  \BibitemOpen
  \bibfield  {author} {\bibinfo {author} {\bibfnamefont {M.}~\bibnamefont {Boehm}}, \bibinfo {author} {\bibfnamefont {S.}~\bibnamefont {Coad}}, \bibinfo {author} {\bibfnamefont {B.}~\bibnamefont {Roessli}}, \bibinfo {author} {\bibfnamefont {A.}~\bibnamefont {Zheludev}}, \bibinfo {author} {\bibfnamefont {M.}~\bibnamefont {Zolliker}}, \bibinfo {author} {\bibfnamefont {P.}~\bibnamefont {Böni}}, \bibinfo {author} {\bibfnamefont {D.~M.}\ \bibnamefont {Paul}}, \bibinfo {author} {\bibfnamefont {H.}~\bibnamefont {Eisaki}}, \bibinfo {author} {\bibfnamefont {N.}~\bibnamefont {Motoyama}},\ and\ \bibinfo {author} {\bibfnamefont {S.}~\bibnamefont {Uchida}},\ }\bibfield  {title} {\bibinfo {title} {Competing exchange interactions in $\mathrm{Li}_{2}\mathrm{CuO}_{2}$},\ }\href {https://doi.org/10.1209/epl/i1998-00322-3} {\bibfield  {journal} {\bibinfo  {journal} {Europhys. Lett.}\ }\textbf {\bibinfo {volume} {43}},\ \bibinfo {pages} {77} (\bibinfo {year} {1998})}\BibitemShut {NoStop}%
\bibitem [{\citenamefont {Mertz}\ \emph {et~al.}(2005)\citenamefont {Mertz}, \citenamefont {Hayn}, \citenamefont {Opahle},\ and\ \citenamefont {Rosner}}]{Mertz2005}%
  \BibitemOpen
  \bibfield  {author} {\bibinfo {author} {\bibfnamefont {D.}~\bibnamefont {Mertz}}, \bibinfo {author} {\bibfnamefont {R.}~\bibnamefont {Hayn}}, \bibinfo {author} {\bibfnamefont {I.}~\bibnamefont {Opahle}},\ and\ \bibinfo {author} {\bibfnamefont {H.}~\bibnamefont {Rosner}},\ }\bibfield  {title} {\bibinfo {title} {Calculated magnetocrystalline anisotropy and magnetic moment distribution in $\mathrm{Li}_{2}\mathrm{CuO}_{2}$},\ }\href {https://doi.org/10.1103/PhysRevB.72.085133} {\bibfield  {journal} {\bibinfo  {journal} {Phys. Rev. B}\ }\textbf {\bibinfo {volume} {72}},\ \bibinfo {pages} {085133} (\bibinfo {year} {2005})}\BibitemShut {NoStop}%
\bibitem [{\citenamefont {Matsuda}\ \emph {et~al.}(1990)\citenamefont {Matsuda}, \citenamefont {Yamada}, \citenamefont {Kakurai}, \citenamefont {Kadowaki}, \citenamefont {Thurston}, \citenamefont {Endoh}, \citenamefont {Hidaka}, \citenamefont {Birgeneau}, \citenamefont {Kastner}, \citenamefont {Gehring}, \citenamefont {Moudden},\ and\ \citenamefont {Shirane}}]{Matsuda1990}%
  \BibitemOpen
  \bibfield  {author} {\bibinfo {author} {\bibfnamefont {M.}~\bibnamefont {Matsuda}}, \bibinfo {author} {\bibfnamefont {K.}~\bibnamefont {Yamada}}, \bibinfo {author} {\bibfnamefont {K.}~\bibnamefont {Kakurai}}, \bibinfo {author} {\bibfnamefont {H.}~\bibnamefont {Kadowaki}}, \bibinfo {author} {\bibfnamefont {T.~R.}\ \bibnamefont {Thurston}}, \bibinfo {author} {\bibfnamefont {Y.}~\bibnamefont {Endoh}}, \bibinfo {author} {\bibfnamefont {Y.}~\bibnamefont {Hidaka}}, \bibinfo {author} {\bibfnamefont {R.~J.}\ \bibnamefont {Birgeneau}}, \bibinfo {author} {\bibfnamefont {M.~A.}\ \bibnamefont {Kastner}}, \bibinfo {author} {\bibfnamefont {P.~M.}\ \bibnamefont {Gehring}}, \bibinfo {author} {\bibfnamefont {A.~H.}\ \bibnamefont {Moudden}},\ and\ \bibinfo {author} {\bibfnamefont {G.}~\bibnamefont {Shirane}},\ }\bibfield  {title} {\bibinfo {title} {Three-dimensional magnetic structures and rare-earth magnetic ordering in $\mathrm{Nd}_{2}\mathrm{CuO}_{4}$ and $\mathrm{Pr}_{2}\mathrm{CuO}_{4}$},\ }\href
  {https://doi.org/10.1103/PhysRevB.42.10098} {\bibfield  {journal} {\bibinfo  {journal} {Phys. Rev. B}\ }\textbf {\bibinfo {volume} {42}},\ \bibinfo {pages} {10098} (\bibinfo {year} {1990})}\BibitemShut {NoStop}%
\bibitem [{\citenamefont {Bourges}\ \emph {et~al.}(1997)\citenamefont {Bourges}, \citenamefont {Casalta}, \citenamefont {Ivanov},\ and\ \citenamefont {Petitgrand}}]{Bourges1997}%
  \BibitemOpen
  \bibfield  {author} {\bibinfo {author} {\bibfnamefont {P.}~\bibnamefont {Bourges}}, \bibinfo {author} {\bibfnamefont {H.}~\bibnamefont {Casalta}}, \bibinfo {author} {\bibfnamefont {A.~S.}\ \bibnamefont {Ivanov}},\ and\ \bibinfo {author} {\bibfnamefont {D.}~\bibnamefont {Petitgrand}},\ }\bibfield  {title} {\bibinfo {title} {Superexchange coupling and spin susceptibility spectral weight in undoped monolayer cuprates},\ }\href {https://doi.org/10.1103/PhysRevLett.79.4906} {\bibfield  {journal} {\bibinfo  {journal} {Phys. Rev. Lett.}\ }\textbf {\bibinfo {volume} {79}},\ \bibinfo {pages} {4906} (\bibinfo {year} {1997})}\BibitemShut {NoStop}%
\bibitem [{\citenamefont {Sumarlin}\ \emph {et~al.}(1995)\citenamefont {Sumarlin}, \citenamefont {Lynn}, \citenamefont {Chattopadhyay}, \citenamefont {Barilo}, \citenamefont {Zhigunov},\ and\ \citenamefont {Peng}}]{Sumarlin1995}%
  \BibitemOpen
  \bibfield  {author} {\bibinfo {author} {\bibfnamefont {I.~W.}\ \bibnamefont {Sumarlin}}, \bibinfo {author} {\bibfnamefont {J.~W.}\ \bibnamefont {Lynn}}, \bibinfo {author} {\bibfnamefont {T.}~\bibnamefont {Chattopadhyay}}, \bibinfo {author} {\bibfnamefont {S.~N.}\ \bibnamefont {Barilo}}, \bibinfo {author} {\bibfnamefont {D.~I.}\ \bibnamefont {Zhigunov}},\ and\ \bibinfo {author} {\bibfnamefont {J.~L.}\ \bibnamefont {Peng}},\ }\bibfield  {title} {\bibinfo {title} {Magnetic structure and spin dynamics of the \textrm{Pr} and \textrm{Cu} in $\mathrm{Pr}_{2}\mathrm{CuO}_{4}$},\ }\href {https://doi.org/10.1103/PhysRevB.51.5824} {\bibfield  {journal} {\bibinfo  {journal} {Phys. Rev. B}\ }\textbf {\bibinfo {volume} {51}},\ \bibinfo {pages} {5824} (\bibinfo {year} {1995})}\BibitemShut {NoStop}%
\bibitem [{\citenamefont {Sulewski}\ \emph {et~al.}(1990)\citenamefont {Sulewski}, \citenamefont {Fleury}, \citenamefont {Lyons}, \citenamefont {Cheong},\ and\ \citenamefont {Fisk}}]{Sulewski}%
  \BibitemOpen
  \bibfield  {author} {\bibinfo {author} {\bibfnamefont {P.~E.}\ \bibnamefont {Sulewski}}, \bibinfo {author} {\bibfnamefont {P.~A.}\ \bibnamefont {Fleury}}, \bibinfo {author} {\bibfnamefont {K.~B.}\ \bibnamefont {Lyons}}, \bibinfo {author} {\bibfnamefont {S.-W.}\ \bibnamefont {Cheong}},\ and\ \bibinfo {author} {\bibfnamefont {Z.}~\bibnamefont {Fisk}},\ }\bibfield  {title} {\bibinfo {title} {Light scattering from quantum spin fluctuations in $\mathrm{R}_{2}\mathrm{CuO}_{4}$ ($\mathrm{R}=\mathrm{La, Nd, Sm}$)},\ }\href {https://doi.org/10.1103/PhysRevB.41.225} {\bibfield  {journal} {\bibinfo  {journal} {Phys. Rev. B}\ }\textbf {\bibinfo {volume} {41}},\ \bibinfo {pages} {225} (\bibinfo {year} {1990})}\BibitemShut {NoStop}%
\bibitem [{\citenamefont {Peng}\ \emph {et~al.}(2017)\citenamefont {Peng}, \citenamefont {Dellea}, \citenamefont {Minola}, \citenamefont {Conni}, \citenamefont {Amorese}, \citenamefont {Di~Castro}, \citenamefont {De~Luca}, \citenamefont {Kummer}, \citenamefont {Salluzzo}, \citenamefont {Sun} \emph {et~al.}}]{peng2017influence}%
  \BibitemOpen
  \bibfield  {author} {\bibinfo {author} {\bibfnamefont {Y.}~\bibnamefont {Peng}}, \bibinfo {author} {\bibfnamefont {G.}~\bibnamefont {Dellea}}, \bibinfo {author} {\bibfnamefont {M.}~\bibnamefont {Minola}}, \bibinfo {author} {\bibfnamefont {M.}~\bibnamefont {Conni}}, \bibinfo {author} {\bibfnamefont {A.}~\bibnamefont {Amorese}}, \bibinfo {author} {\bibfnamefont {D.}~\bibnamefont {Di~Castro}}, \bibinfo {author} {\bibfnamefont {G.}~\bibnamefont {De~Luca}}, \bibinfo {author} {\bibfnamefont {K.}~\bibnamefont {Kummer}}, \bibinfo {author} {\bibfnamefont {M.}~\bibnamefont {Salluzzo}}, \bibinfo {author} {\bibfnamefont {X.}~\bibnamefont {Sun}}, \emph {et~al.},\ }\bibfield  {title} {\bibinfo {title} {Influence of apical oxygen on the extent of in-plane exchange interaction in cuprate superconductors},\ }\href {https://doi.org/https://doi.org/10.1038/nphys4248} {\bibfield  {journal} {\bibinfo  {journal} {Nat. Phys.}\ }\textbf {\bibinfo {volume} {13}},\ \bibinfo {pages} {1201} (\bibinfo {year} {2017})}\BibitemShut
  {NoStop}%
\bibitem [{\citenamefont {Kotegawa}\ \emph {et~al.}(2004)\citenamefont {Kotegawa}, \citenamefont {Tokunaga}, \citenamefont {Araki}, \citenamefont {Zheng}, \citenamefont {Kitaoka}, \citenamefont {Tokiwa}, \citenamefont {Ito}, \citenamefont {Watanabe}, \citenamefont {Iyo}, \citenamefont {Tanaka},\ and\ \citenamefont {Ihara}}]{Kotegawa2004}%
  \BibitemOpen
  \bibfield  {author} {\bibinfo {author} {\bibfnamefont {H.}~\bibnamefont {Kotegawa}}, \bibinfo {author} {\bibfnamefont {Y.}~\bibnamefont {Tokunaga}}, \bibinfo {author} {\bibfnamefont {Y.}~\bibnamefont {Araki}}, \bibinfo {author} {\bibfnamefont {G.-q.}\ \bibnamefont {Zheng}}, \bibinfo {author} {\bibfnamefont {Y.}~\bibnamefont {Kitaoka}}, \bibinfo {author} {\bibfnamefont {K.}~\bibnamefont {Tokiwa}}, \bibinfo {author} {\bibfnamefont {K.}~\bibnamefont {Ito}}, \bibinfo {author} {\bibfnamefont {T.}~\bibnamefont {Watanabe}}, \bibinfo {author} {\bibfnamefont {A.}~\bibnamefont {Iyo}}, \bibinfo {author} {\bibfnamefont {Y.}~\bibnamefont {Tanaka}},\ and\ \bibinfo {author} {\bibfnamefont {H.}~\bibnamefont {Ihara}},\ }\bibfield  {title} {\bibinfo {title} {Coexistence of superconductivity and antiferromagnetism in multilayered high-${T}_{c}$ superconductor $\mathrm{HgBa}_{2}\mathrm{Ca}_{4}\mathrm{Cu}_{5}\mathrm{O}_{y}:$ \textrm{Cu-NMR} study},\ }\href {https://doi.org/10.1103/PhysRevB.69.014501} {\bibfield  {journal}
  {\bibinfo  {journal} {Phys. Rev. B}\ }\textbf {\bibinfo {volume} {69}},\ \bibinfo {pages} {014501} (\bibinfo {year} {2004})}\BibitemShut {NoStop}%
\bibitem [{\citenamefont {Mukuda}\ \emph {et~al.}(2006)\citenamefont {Mukuda}, \citenamefont {Abe}, \citenamefont {Araki}, \citenamefont {Kitaoka}, \citenamefont {Tokiwa}, \citenamefont {Watanabe}, \citenamefont {Iyo}, \citenamefont {Kito},\ and\ \citenamefont {Tanaka}}]{Mukuda2006}%
  \BibitemOpen
  \bibfield  {author} {\bibinfo {author} {\bibfnamefont {H.}~\bibnamefont {Mukuda}}, \bibinfo {author} {\bibfnamefont {M.}~\bibnamefont {Abe}}, \bibinfo {author} {\bibfnamefont {Y.}~\bibnamefont {Araki}}, \bibinfo {author} {\bibfnamefont {Y.}~\bibnamefont {Kitaoka}}, \bibinfo {author} {\bibfnamefont {K.}~\bibnamefont {Tokiwa}}, \bibinfo {author} {\bibfnamefont {T.}~\bibnamefont {Watanabe}}, \bibinfo {author} {\bibfnamefont {A.}~\bibnamefont {Iyo}}, \bibinfo {author} {\bibfnamefont {H.}~\bibnamefont {Kito}},\ and\ \bibinfo {author} {\bibfnamefont {Y.}~\bibnamefont {Tanaka}},\ }\bibfield  {title} {\bibinfo {title} {Uniform mixing of high-${T}_{c}$ superconductivity and antiferromagnetism on a single $\mathrm{CuO}_{2}$ plane of a $\mathrm{Hg}$-based five-layered cuprate},\ }\href {https://doi.org/10.1103/PhysRevLett.96.087001} {\bibfield  {journal} {\bibinfo  {journal} {Phys. Rev. Lett.}\ }\textbf {\bibinfo {volume} {96}},\ \bibinfo {pages} {087001} (\bibinfo {year} {2006})}\BibitemShut {NoStop}%
\bibitem [{\citenamefont {Plakida}(2010)}]{Plakida_2010}%
  \BibitemOpen
  \bibfield  {author} {\bibinfo {author} {\bibfnamefont {N.}~\bibnamefont {Plakida}},\ }\href {https://doi.org/10.1007/978-3-642-12633-8} {\emph {\bibinfo {title} {High-Temperature Cuprate Superconductors}}}\ (\bibinfo  {publisher} {Springer Berlin Heidelberg},\ \bibinfo {year} {2010})\BibitemShut {NoStop}%
\bibitem [{\citenamefont {Liu}\ \emph {et~al.}(2005)\citenamefont {Liu}, \citenamefont {Wen}, \citenamefont {Shan}, \citenamefont {Yang}, \citenamefont {Lu}, \citenamefont {Gao}, \citenamefont {Park}, \citenamefont {Jung},\ and\ \citenamefont {Lee}}]{Liu_2005}%
  \BibitemOpen
  \bibfield  {author} {\bibinfo {author} {\bibfnamefont {Z.~Y.}\ \bibnamefont {Liu}}, \bibinfo {author} {\bibfnamefont {H.~H.}\ \bibnamefont {Wen}}, \bibinfo {author} {\bibfnamefont {L.}~\bibnamefont {Shan}}, \bibinfo {author} {\bibfnamefont {H.~P.}\ \bibnamefont {Yang}}, \bibinfo {author} {\bibfnamefont {X.~F.}\ \bibnamefont {Lu}}, \bibinfo {author} {\bibfnamefont {H.}~\bibnamefont {Gao}}, \bibinfo {author} {\bibfnamefont {M.-S.}\ \bibnamefont {Park}}, \bibinfo {author} {\bibfnamefont {C.~U.}\ \bibnamefont {Jung}},\ and\ \bibinfo {author} {\bibfnamefont {S.-I.}\ \bibnamefont {Lee}},\ }\bibfield  {title} {\bibinfo {title} {Bulk evidence for s-wave pairing symmetry in electron-doped infinite-layer cuprate $\mathrm{Sr}_{0.9}\mathrm{La}_{0.1}\mathrm{CuO}_2$},\ }\href {https://doi.org/10.1209/epl/i2003-10315-8} {\bibfield  {journal} {\bibinfo  {journal} {Europhys. Lett.}\ }\textbf {\bibinfo {volume} {69}},\ \bibinfo {pages} {263} (\bibinfo {year} {2005})}\BibitemShut {NoStop}%
\bibitem [{\citenamefont {Alloul}\ \emph {et~al.}(2009)\citenamefont {Alloul}, \citenamefont {Bobroff}, \citenamefont {Gabay},\ and\ \citenamefont {Hirschfeld}}]{Alloul2009}%
  \BibitemOpen
  \bibfield  {author} {\bibinfo {author} {\bibfnamefont {H.}~\bibnamefont {Alloul}}, \bibinfo {author} {\bibfnamefont {J.}~\bibnamefont {Bobroff}}, \bibinfo {author} {\bibfnamefont {M.}~\bibnamefont {Gabay}},\ and\ \bibinfo {author} {\bibfnamefont {P.~J.}\ \bibnamefont {Hirschfeld}},\ }\bibfield  {title} {\bibinfo {title} {Defects in correlated metals and superconductors},\ }\href {https://doi.org/10.1103/RevModPhys.81.45} {\bibfield  {journal} {\bibinfo  {journal} {Rev. Mod. Phys.}\ }\textbf {\bibinfo {volume} {81}},\ \bibinfo {pages} {45} (\bibinfo {year} {2009})}\BibitemShut {NoStop}%
\bibitem [{\citenamefont {Christensen}\ \emph {et~al.}(2011)\citenamefont {Christensen}, \citenamefont {Hirschfeld},\ and\ \citenamefont {Andersen}}]{Christensen2011}%
  \BibitemOpen
  \bibfield  {author} {\bibinfo {author} {\bibfnamefont {R.~B.}\ \bibnamefont {Christensen}}, \bibinfo {author} {\bibfnamefont {P.~J.}\ \bibnamefont {Hirschfeld}},\ and\ \bibinfo {author} {\bibfnamefont {B.~M.}\ \bibnamefont {Andersen}},\ }\bibfield  {title} {\bibinfo {title} {Two routes to magnetic order by disorder in underdoped cuprates},\ }\href {https://doi.org/10.1103/PhysRevB.84.184511} {\bibfield  {journal} {\bibinfo  {journal} {Phys. Rev. B}\ }\textbf {\bibinfo {volume} {84}},\ \bibinfo {pages} {184511} (\bibinfo {year} {2011})}\BibitemShut {NoStop}%
\bibitem [{\citenamefont {Murayama}\ \emph {et~al.}(1989)\citenamefont {Murayama}, \citenamefont {Mori}, \citenamefont {Yomo}, \citenamefont {Takagi}, \citenamefont {Uchida},\ and\ \citenamefont {Tokura}}]{murayama1989anomalous}%
  \BibitemOpen
  \bibfield  {author} {\bibinfo {author} {\bibfnamefont {C.}~\bibnamefont {Murayama}}, \bibinfo {author} {\bibfnamefont {N.}~\bibnamefont {Mori}}, \bibinfo {author} {\bibfnamefont {S.}~\bibnamefont {Yomo}}, \bibinfo {author} {\bibfnamefont {H.}~\bibnamefont {Takagi}}, \bibinfo {author} {\bibfnamefont {S.}~\bibnamefont {Uchida}},\ and\ \bibinfo {author} {\bibfnamefont {Y.}~\bibnamefont {Tokura}},\ }\bibfield  {title} {\bibinfo {title} {Anomalous absence of pressure effect on transition temperature in the electron-doped superconductor $\mathrm{Nd}_{1.85}\mathrm{Ce}_{0.15}\mathrm{CuO}_{4-\delta}$},\ }\href {https://doi.org/https://doi.org/10.1038/339293a0} {\bibfield  {journal} {\bibinfo  {journal} {Nature}\ }\textbf {\bibinfo {volume} {339}},\ \bibinfo {pages} {293} (\bibinfo {year} {1989})}\BibitemShut {NoStop}%
\bibitem [{\citenamefont {Ishiwata}\ \emph {et~al.}(2013)\citenamefont {Ishiwata}, \citenamefont {Kotajima}, \citenamefont {Takeshita}, \citenamefont {Terakura}, \citenamefont {Seki},\ and\ \citenamefont {Tokura}}]{Ishiwata2013}%
  \BibitemOpen
  \bibfield  {author} {\bibinfo {author} {\bibfnamefont {S.}~\bibnamefont {Ishiwata}}, \bibinfo {author} {\bibfnamefont {D.}~\bibnamefont {Kotajima}}, \bibinfo {author} {\bibfnamefont {N.}~\bibnamefont {Takeshita}}, \bibinfo {author} {\bibfnamefont {C.}~\bibnamefont {Terakura}}, \bibinfo {author} {\bibfnamefont {S.}~\bibnamefont {Seki}},\ and\ \bibinfo {author} {\bibfnamefont {Y.}~\bibnamefont {Tokura}},\ }\bibfield  {title} {\bibinfo {title} {Optimal $\mathit{T}_c$ for electron-doped cuprate realized under high pressure},\ }\href {https://doi.org/10.7566/JPSJ.82.063705} {\bibfield  {journal} {\bibinfo  {journal} {J. Phys. Soc. of Jpn}\ }\textbf {\bibinfo {volume} {82}},\ \bibinfo {pages} {063705} (\bibinfo {year} {2013})}\BibitemShut {NoStop}%
\bibitem [{\citenamefont {Takahashi}\ \emph {et~al.}(1994)\citenamefont {Takahashi}, \citenamefont {Môri}, \citenamefont {Azuma}, \citenamefont {Hiroi},\ and\ \citenamefont {Takano}}]{TAKAHASHI1994395}%
  \BibitemOpen
  \bibfield  {author} {\bibinfo {author} {\bibfnamefont {H.}~\bibnamefont {Takahashi}}, \bibinfo {author} {\bibfnamefont {N.}~\bibnamefont {Môri}}, \bibinfo {author} {\bibfnamefont {M.}~\bibnamefont {Azuma}}, \bibinfo {author} {\bibfnamefont {Z.}~\bibnamefont {Hiroi}},\ and\ \bibinfo {author} {\bibfnamefont {M.}~\bibnamefont {Takano}},\ }\bibfield  {title} {\bibinfo {title} {{Effect of pressure on T$_c$ of hole- and electron-doped infinite-layer compounds up to 8 \textrm{GPa}}},\ }\href {https://doi.org/https://doi.org/10.1016/0921-4534(94)90099-X} {\bibfield  {journal} {\bibinfo  {journal} {Physica C}\ }\textbf {\bibinfo {volume} {227}},\ \bibinfo {pages} {395} (\bibinfo {year} {1994})}\BibitemShut {NoStop}%
\bibitem [{\citenamefont {Castro}\ \emph {et~al.}(2009)\citenamefont {Castro}, \citenamefont {Khasanov}, \citenamefont {Shengelaya}, \citenamefont {Conder}, \citenamefont {Jang}, \citenamefont {Park}, \citenamefont {Lee},\ and\ \citenamefont {Keller}}]{DiCastro_2009}%
  \BibitemOpen
  \bibfield  {author} {\bibinfo {author} {\bibfnamefont {D.~D.}\ \bibnamefont {Castro}}, \bibinfo {author} {\bibfnamefont {R.}~\bibnamefont {Khasanov}}, \bibinfo {author} {\bibfnamefont {A.}~\bibnamefont {Shengelaya}}, \bibinfo {author} {\bibfnamefont {K.}~\bibnamefont {Conder}}, \bibinfo {author} {\bibfnamefont {D.-J.}\ \bibnamefont {Jang}}, \bibinfo {author} {\bibfnamefont {M.-S.}\ \bibnamefont {Park}}, \bibinfo {author} {\bibfnamefont {S.-I.}\ \bibnamefont {Lee}},\ and\ \bibinfo {author} {\bibfnamefont {H.}~\bibnamefont {Keller}},\ }\bibfield  {title} {\bibinfo {title} {Comparative study of the pressure effects on the magnetic penetration depth in electron- and hole-doped cuprate superconductors},\ }\href {https://doi.org/10.1088/0953-8984/21/27/275701} {\bibfield  {journal} {\bibinfo  {journal} {J. of Phys.: Condens. Matter}\ }\textbf {\bibinfo {volume} {21}},\ \bibinfo {pages} {275701} (\bibinfo {year} {2009})}\BibitemShut {NoStop}%
\bibitem [{\citenamefont {Kim}\ \emph {et~al.}(2006)\citenamefont {Kim}, \citenamefont {Kim}, \citenamefont {Jung}, \citenamefont {Park},\ and\ \citenamefont {Lee}}]{KIM2006}%
  \BibitemOpen
  \bibfield  {author} {\bibinfo {author} {\bibfnamefont {H.}~\bibnamefont {Kim}}, \bibinfo {author} {\bibfnamefont {M.-H.}\ \bibnamefont {Kim}}, \bibinfo {author} {\bibfnamefont {M.}~\bibnamefont {Jung}}, \bibinfo {author} {\bibfnamefont {M.-S.}\ \bibnamefont {Park}},\ and\ \bibinfo {author} {\bibfnamefont {S.-I.}\ \bibnamefont {Lee}},\ }\bibfield  {title} {\bibinfo {title} {Pressure effect on the infinite-layer superconductor $\mathrm{Sr}_{0.9}\mathrm{La}_{0.1}\mathrm{CuO}_2$ studied by magnetization},\ }\href {https://doi.org/https://doi.org/10.1016/j.physb.2006.01.325} {\bibfield  {journal} {\bibinfo  {journal} {Physica B}\ }\textbf {\bibinfo {volume} {378-380}},\ \bibinfo {pages} {886} (\bibinfo {year} {2006})},\ \bibinfo {note} {{Proceedings of the International Conference on Strongly Correlated Electron Systems}}\BibitemShut {NoStop}%
\bibitem [{\citenamefont {Kund}\ \emph {et~al.}(1998)\citenamefont {Kund}, \citenamefont {Neumeier}, \citenamefont {Andres}, \citenamefont {Markl},\ and\ \citenamefont {Saemann-Ischenko}}]{KUND1998173}%
  \BibitemOpen
  \bibfield  {author} {\bibinfo {author} {\bibfnamefont {M.}~\bibnamefont {Kund}}, \bibinfo {author} {\bibfnamefont {J.}~\bibnamefont {Neumeier}}, \bibinfo {author} {\bibfnamefont {K.}~\bibnamefont {Andres}}, \bibinfo {author} {\bibfnamefont {J.}~\bibnamefont {Markl}},\ and\ \bibinfo {author} {\bibfnamefont {G.}~\bibnamefont {Saemann-Ischenko}},\ }\bibfield  {title} {\bibinfo {title} {Large anisotropic pressure effects in electron-doped $\mathrm{Sm}_{2-x}\mathrm{Ce}_x\mathrm{CuO}_{4-y}$},\ }\href {https://doi.org/https://doi.org/10.1016/S0921-4534(97)01842-X} {\bibfield  {journal} {\bibinfo  {journal} {Physica C}\ }\textbf {\bibinfo {volume} {296}},\ \bibinfo {pages} {173} (\bibinfo {year} {1998})}\BibitemShut {NoStop}%
\bibitem [{\citenamefont {Kaga}\ \emph {et~al.}(2005)\citenamefont {Kaga}, \citenamefont {Sasagawa}, \citenamefont {Takahashi}, \citenamefont {Unosawa},\ and\ \citenamefont {Takagi}}]{KAGA2005442}%
  \BibitemOpen
  \bibfield  {author} {\bibinfo {author} {\bibfnamefont {Y.}~\bibnamefont {Kaga}}, \bibinfo {author} {\bibfnamefont {T.}~\bibnamefont {Sasagawa}}, \bibinfo {author} {\bibfnamefont {S.}~\bibnamefont {Takahashi}}, \bibinfo {author} {\bibfnamefont {K.}~\bibnamefont {Unosawa}},\ and\ \bibinfo {author} {\bibfnamefont {H.}~\bibnamefont {Takagi}},\ }\bibfield  {title} {\bibinfo {title} {Uniaxial pressure effect in electron-doped high-temperature superconductor $\mathrm{Nd}_{1.84}\mathrm{Ce}_{0.16}\mathrm{CuO}_4$},\ }\href {https://doi.org/https://doi.org/10.1016/j.physb.2005.01.091} {\bibfield  {journal} {\bibinfo  {journal} {Physica B}\ }\textbf {\bibinfo {volume} {359-361}},\ \bibinfo {pages} {442} (\bibinfo {year} {2005})}\BibitemShut {NoStop}%
\end{thebibliography}%

\end{document}